\documentclass[11pt]{article}
\usepackage[all]{xy}
\usepackage{amsmath,amsfonts,amssymb,amsthm,amscd,pb-diagram,pb-xy}
\usepackage{graphicx}
\usepackage{epsfig}
\usepackage{caption}

\newcounter{ExCounter}
\stepcounter{ExCounter}

\newcounter{RemCounter}
\stepcounter{RemCounter}

\newenvironment{Remark}{\par\medskip\noindent{\sc Remark \arabic{RemCounter}\stepcounter{RemCounter}: \, }}{\hfill $\Diamond$\par\medskip}
\newenvironment{Example}{\par\medskip\noindent{\sc Example \arabic{ExCounter}\stepcounter{ExCounter}: \, }}{\hfill $\triangle$\par\medskip}

\newtheorem{Theorem}{Theorem}[section]
\newtheorem{Definition}{Definition}[section]

\newcommand{\Reali}{\ensuremath{\mathbb{R}}}
\newcommand{\Euclideo}{\ensuremath{\mathbb{E}}}

\def\={\, = \,}
\def\vth{\vartheta}
\def\vph{\varphi}
\def\A{\mathcal A\/}
\def\B{\mathcal B\/}

\def\H{\mathcal H\/}
\def\K{\mathcal K\/}
\def\L{\mathcal L\/}
\def\M{\mathcal M\/}
\def\N{\mathcal N\/}

\def\Q{\mathcal Q\/}
\def\P{\mathcal P\/}
\def\R{\mathcal R\/}
\def\S{\mathcal S\/}
\def\PJ{\mathcal P_{\S}\/}
\def\PJO{\mathcal P^{\perp}_{\S}\/}
\def\PA{\mathcal P^{\|}_{\A}\/}
\def\PAO{\mathcal P^{\perp}_{\A}\/}

\def\PV{\mathcal P^{\|}\/}
\def\PVO{\mathcal P^{\perp}\/}

\def\S{\mathcal S\/}

\def\J{J_1\/(\M)}
\def\JJ{J_2\/(\M)}
\def\JS{J_1\/(\S)}
\def\IJS{(i_*)^*(J_1\/(\M))}

\def\iJS{i_*(J_1\/(\S))}
\def\iV{i_*(V\/(\S))}

\def\vel{{\bf V}}
\def\velort{{\bf V}^{\perp}}
\def\velortS{{\bf V}^{\perp}_{\S}\/}
\def\velortA{{\bf V}^{\perp}_{\A}\/}
\def\velortB{{\bf V}^{\perp}_{\B}\/}

\begin{document}

\title{A survey about framing the bases of Impulsive Mechanics of constrained systems into a jet--bundle geometric context}

\author{Stefano Pasquero \\
            Department of Mathematical, Physical and Computer Sciences \\
            University of Parma \\
            Parco Area delle Scienze 53/a (Campus) \\
            43124 PARMA \\
            email: stefano.pasquero@unipr.it}

\date{\today}

\maketitle

\begin{abstract}
\noindent


We illustrate how the different kinds of constraints acting on an
impulsive mechanical system can be clearly described in the
geometric setup given by the configuration space--time bundle
$\pi_t:\M \to \Euclideo$ and its first jet extension $\pi: \J \to
\M$ in a way that ensures total compliance with axioms and
invariance requirements of Classical Mechanics. We specify the
differences between geometric and constitutive characterizations
of a constraint. We point out the relevance of the role played by
the concept of frame of reference, underlining when the frame
independence is mandatorily required and when a choice of a frame
is an inescapable need. The thorough rationalization allows the
introduction of unusual but meaningful kinds of constraints, such
as unilateral kinetic constraints or breakable constraints, and of
new theoretical aspects, such as the possible dependence of the
impulsive reaction by the active forces acting on the system.

\medskip


\noindent {\bf PACS2010:} 45.50.Tn -- 45.20.D- -- 45.10.Na

\medskip


\end{abstract}

\section*{Introduction}

A modern approach to the study of Classical Mechanics of systems
with a finite number of degrees of freedom cannot prescind from
structures and methods of Differential Geometry, mainly in order
to construct a powerful setup to frame the theory with its
invariance requirements, and from methods and results of
Mathematical Analysis, mainly in order to solve and analyze the
equations governing the motion. However, of course, Classical
Mechanics does not consist only of Differential Geometry and
Mathematical Analysis. It has some basic concepts strictly of its
own, such as causality, determinism, coordinate description
invariance and frame invariance, that must be mandatorily taken
into account in the construction of a mechanical theory.

The theory of jet--extensions of fibred manifolds (see, e.g.
\cite{Saunders,Pommaret}) is, at the present time, the best common
ground where both the invariant description usually provided by a
differential geometric setup and the powerful techniques provided
by Mathematical Analysis about time dependent ordinary
differential equations can be synergically applied to study
Classical Mechanics.

Unfortunately this synergy is not so effective for the study of
Classical Impulsive Mechanics. In fact the impulsive aspects of
the phenomenon, with their intrinsic discontinuity of the
velocities, suggest, regarding the analytical methods, the use of
concepts like measure differential equations, bounded values and
absolutely continuous functions and the related techniques. These
are usually framed in an $\Reali^n$ environment, but they do not
have yet a well founded basis on generic differentiable manifolds.
The standard artifice of using the local description of a
riemannian manifold $\Q$ or of a product bundle $\Reali \times \Q$
taken as configuration space or space--time of the mechanical
system is not a way around the problem of giving a meaningful
concept of measure differential equations satisfying local
coordinate invariance. Moreover the introduction of $\Q$ or
$\Reali \times \Q$ is structurally unfit to take into account the
frame invariance of the mechanical theory. The goal of a
coordinate and frame invariance of the approach to a mechanical
theory can for now be achieved only using a space--time bundle
$\pi_t: \M \to \Euclideo$ (being $\Euclideo$ the euclidean
$1$--dimensional line) as configuration space--time of the system,
but in this context, due to its natural lack of a riemannian
structure, an invariant theory of measure differential equations
is actually not developed.

Then, until a well founded analytical theory of measure
differential equations will be constructed in a jet--bundle
context, the study of Classical Impulsive Mechanics presents two
alternative approaches: the analytical one, powered by all the
techniques of Mathematical Analysis but weakened by the
difficulties in having a coordinate and/or frame invariant
character, and the geometric one that, thanks to the powerful
techniques of generic coordinates and vectors fields, satisfies
the invariance requirements of a mechanical theory but that can
menage the equations of motion with great difficulty.

A remarkable exception to this difficulty in the case of geometric
approach consists in the so--called ``{\it event--driven}''
algebraic approach to Impulsive Mechanics. In fact it gives rise
to finite, and not differential, evolution equations, so that, in
this context, the importance of the techniques of Mathematical
Analysis is partly downsized.

\bigskip

In this paper we advocate and show that the geometric environment
determined by the first order jet--bundle $\J$  of the
configuration space--time bundle $\pi_t: \M \to \Euclideo$
together with some suitable subbundles combines three important
qualities: first, the peculiar requirements of a mechanical theory
(causality, determinism, coordinate and frame invariance) can be
correctly introduced; second, every result pertaining the ``first
order'' aspect of the theory (that is, results pertaining position
and velocity of the system) can be framed in this first order
jet--bundle context; third, when we restrict our attention to the
event--driven algebraic approach, it gives powerful techniques to
model, to perform a deep analysis and to achieve results regarding
the impulsive phenomenon. These qualities hold both for free
impulsive mechanical systems and also, especially, for constrained
impulsive mechanical systems and in particular for systems with a
finite number of degrees of freedom subject to additional
unilateral constraints.

This paper has then several aims: to provide a clarifying
organization of the operational and applicable bases of Impulsive
Mechanics of (constrained) systems with a finite number of degrees
of freedom consistent with the axioms and principles of Classical
Mechanics; to gather in a single work the majority of known ideas
about the jet--bundle approach to these basic aspects of Impulsive
Mechanics, with a detailed analysis of the possible additional
impulsive constraints acting on the system; to point out the
crudity, the inaccuracy, or even the inconsistency of a geometric
framework for Impulsive Mechanics based on manifolds that are not
fibred or on product manifolds; to suggest new perspectives for
possible future investigations.

\bigskip

The very brief Preliminaries section contains the heuristic
descriptions of the impulsive mechanical problem and of the range
of applicability of the impulsive approach to the evolution of a
mechanical system. The formal aspect of the theory are not taken
into account in this section and will be described in the
following ones.

Section 1 has an introductory character: we recall the geometric
setup for a time--dependent and frame--independent description of
the behavior of an impulsive free mechanical system. Since we are
not interested in a force--acceleration description of this
behavior, but only in an impulse--velocity description, we limit
our presentation to the geometry of the configuration space--time
bundle $\M \to \Euclideo$ and the absolute velocity space--time
bundle $\J \to \M$ of the system. We show that this geometric
context forms a very natural environment where describing
impulsive behaviors, moreover underlining the correct causality of
the formulation of the Integrated, or Impulsive, Newton Law
(briefly INL) as presented, for example, in
\cite{LeviCivita,Pars,Peres,Stronge}. Furthermore, we show that
the commonly used geometric framework given by the simple
configuration space $\Q$, or by the cartesian products $\Reali
\times \Q$ or $\Euclideo \times \Q$ is not fit for a frame
independent description of the behavior of the system, since it
implies the use of an intrinsically defined frame of reference.

Section 2 concerns the geometric setup  for a time--dependent and
frame--independent description of impulsive mechanical systems
subject to constraints. We show that we can embody the wide
variety of constraints possibly acting on the system in a single
geometric structure formed by suitable subbundles of the geometric
setup describing a free system. Using this structure, we can
classify the constraints on the basis of their geometric
properties: positional or kinetic, bilateral or unilateral,
permanent or instantaneous, single or multiple, isotropic or
anisotropic.

It is well known that the geometric properties are not sufficient
to classify the whole kinds of constraints, since they can be
classified also depending on the nature of their action on the
impulsive system. Section 3 concerns the constitutive
possibilities of the various kinds of constraints described in the
previous section. After a brief reminder of the key role played by
the concept of constitutive characterization of a constraint, we
distinguish the constraints on the basis of the nature of their
constitutive characterization: ideal or non--ideal, with or
without friction, breakable or unbreakable.

Section 4 concludes the paper presenting a condensed summary of
rationalizations, ideas and clarifications, some of them having an
innovative character, distributed along the paper.

The list of possible references about impulsive constrained
systems is very huge, and a bibliography claiming to be exhaustive
on the argument should be excessively long compared to the length
of the paper. Moreover, only few works would be reasonably
pertinent to the specific approach presented in the paper and a
long list would then draw away the attention of the reader from
the peculiarities (especially about the foundational aspects such
as the respect of axioms and principles of Classical Mechanics,
coordinate and frame invariance) of this paper. Therefore, the
list of references has been based on the minimality criterion of
making the paper self--consistent. Anyway, for large but not
recent or exhaustive lists of general references, see for example
the books
\cite{Brogliato2000impacts,Pfeiffer2000multibody,Johnson,Stronge}.

Several remarks and examples are presented in the paper. In order
to improve the typographical clearness, the signs $\Diamond$ and
$\triangle$ are placed at the end of remarks and examples
respectively.

\section*{Preliminaries}

An impulsive behavior in a ``single point'' of a mechanical system
with a finite number of degrees of freedom is a time--evolution of
the system such that the map assigning the position of the system
is continuous for every instant while the map assigning the
velocity of the system is continuous in all except a single
instant, and in this instant of discontinuity the velocities are
subject to a finite jump.

This manifestly heuristic definition will be clearer once the
correct geometric setup of the problem will be described. It is
however clear that no real mechanical system has such a behavior,
and that as a matter of principle the correct description of such
phenomenon should involve a ``very small'' but not singular time
interval in which the evolution should be studied using
deformation, elasticity, thermodynamics, acoustics and so on. Then
the definition is a purely theoretical model, a limit situation of
motions of  systems where some changes of velocity are sudden
enough to be considered instantaneous, or at least such that we
are not interested in a detailed knowledge (or we are not able to
perform a precise analysis) of what happens to the system between
two ``very near'' instants.

However, overlooking this evident formal vulnerability, the
examples of evolutions of mechanical systems for which such an
approximation provides a useful context of analysis are numerous
and physically meaningful: from the completely theoretical problem
of a free system subject to a single active impulse to the wide
records of possible impacts happening in a billiard situation
(single impact of type ball--cushion, multiple impact of type
ball--cushions, single impact ball--ball, multiple impact
ball--balls and so on, all of them with or without friction, with
or without pure rolling conditions, and so on), from the multiple
impact problem inherent in the motion of a granular material to
the single impact problem of a steel sphere on a bulletproof
glass.

\medskip

Another critical point of the previous heuristic definition of
impulsive behavior is the uniqueness of the point of discontinuity
of the velocity. This is an only apparent weak point, since by its
very nature every time evolution of a mechanical system can be
studied locally (with respect to time) and every impulsive
behavior of a system can happen only in an isolated instant,
forerun and followed by two non singular time intervals where the
maps of position and velocity are both continuous. This justifies
the so called event--driven approach, where the regular motion of
the system just before the instant of discontinuity determines the
initial conditions of the impulsive problem, whose ``solution''
determines the initial conditions of the regular motion of the
system just after the discontinuity.

Note that the local character of the impulsive problem implies
that we can neglect the local or global character of the geometric
objects and of the properties we will introduce in the paper.
Nevertheless, the possibility of restricting our attention to a
local description of the behavior of the system must not misdirect
our attention from the effective mechanical aspects of the theory,
in particular the mandatory respect of axioms and principles of
Classical Mechanics.

\section{Free Systems}

In this section we describe the geometric setup suited to study
Impulsive Mechanics of a free system, and we introduce the
impulsive problem in the geometric context.

\subsection{Geometry of Free systems} The {\it configuration space--time}
 of a mechanical system with a finite number $n$
of degrees of freedom is a fiber bundle $\pi_t: \M \to \Euclideo$
where $\M$ is a $(n+1)$-dimensional differentiable manifold and
$\Euclideo$ is the $1$-dimensional Euclidean space. The fibers of
the bundle $\M$ are diffeomorphic to an $n$-dimensional manifold
$\Q$, usually called the {\it configuration space} of the system.
Then $\M$ is at least locally (and possibly globally)
diffeomorphic to the product bundle $\Reali \times \Q$.

If $t$ is a global cartesian coordinate on $\Euclideo$, the
projection $t\circ\pi_t: \M \to \Reali$ represents the embodiment
of the Absolute Time Axiom in this context. The fiber
$(t\circ\pi_t)^{-1}(t_0)$, diffeomorphic to $\Q$, is the
configuration space of the system at the instant $t_0$.

The absolute time coordinate $t$ and the fibred structure of $\M$
select the family of {\it admissible coordinates} $(t, x^1,
\ldots, x^n)$ having $t$ as first coordinate, and the subgroup
\begin{eqnarray}\label{VarCoordM}
\left\{
\begin{array}{lcl}
\bar{t} &\=& t \, + \, const\\
\bar{x}^i & \= & \bar{x}^i(t, x^1, \ldots, x^n)
\end{array}
\right.
\end{eqnarray}
of the group of coordinate transformations leaving invariant the
$1$--form $dt$ (that is, the measurements of time intervals) and
the fibres of $\M$ (that is, the configuration spaces at the
various instants).

\begin{Remark} The coordinate transformation group (\ref{VarCoordM}) is
different from the coordinate transformation groups
\begin{eqnarray*}
\left\{
\begin{array}{lcl}
\bar{t} &\=& \bar{t}(t) \\
\bar{x}^i & \= & \bar{x}^i(x^1, \ldots, x^n)
\end{array}
\right. \qquad {\rm or} \qquad \left\{
\begin{array}{lcl}
\bar{t} &\=& t \\
\bar{x}^i & \= & \bar{x}^i(x^1, \ldots, x^n)
\end{array}
\right.
\end{eqnarray*}
respecting only the product structures of $\Euclideo \times \Q$
and $\Reali \times \Q$ respectively. \end{Remark}

A motion of the system is a section  $\gamma:\Euclideo \to \M$,
locally represented by a map $t \rightsquigarrow (t, x^1(t),
\ldots, x^n(t))$.

\medskip

The first jet--extension $\pi: \J \to \M$ of the bundle $\M$ is
the {\it absolute velocity space--time} of the system. It is the
$(2n+1)$--dimensional affine subbundle of the tangent bundle
$T(\M)$ described by the condition
\begin{equation*}
\J \= \{ {\bf X} \in T(\M) \, | \, <{\bf X} , \pi_t^*(dt)> =1 \} ,
\end{equation*}
that is the set of vectors tangent to any possible motion of the
system in any point. Using admissible coordinates, the elements of
$\J$ have the form ${\bf p} \= \frac{\partial}{\partial t} + p^1
\frac{\partial}{\partial x^1}+ \ldots + p^n
\frac{\partial}{\partial x^n}$, and they are also called {\it
time--like vectors}. $\J$ can be locally described by
jet--coordinates $(t, x^i,\dot{x}^i), \, i=1,\ldots,n$.

\medskip

The affine jet--bundle $\J$ is modelled on the
$(2n+1)$--dimensional vector bundle
\begin{equation*}
V(\M) \= \{ {\bf X} \in T(\M) \, | \, <{\bf X} , \pi_t^*(dt)> =0
\}
\end{equation*}
of the vertical vectors of $T(\M)$, that is the vectors that are
tangent to the fiber of $\M$. Using admissible coordinates, the
elements of $V(\M)$ have the form ${\bf V} \= V^1
\frac{\partial}{\partial x^1}+ \ldots + V^n
\frac{\partial}{\partial x^n}$, and they are also called {\it
space--like vectors}. $V(\M)$ too can be locally described by the
coordinates $(t, x^i,\dot{x}^i), \, i=1,\ldots,n$. The (fibred)
action of $V(\M)$ on $\J$ is the obvious fibred sum
\begin{eqnarray}\label{ActionVJ}
\begin{array}{rcccl}
+ : & \J \times V(\M) & \to & \J &\quad \mbox{s.t.}\\
& ({\bf p} , {\bf V}) & \rightsquigarrow & {\bf p} + {\bf V} &\quad  \Leftrightarrow \\
& ((t, x^i,p^i), (t, x^i,V^i)) & \rightsquigarrow & (t, x^i,p^i +
V^i) .
\end{array}
\end{eqnarray}

\medskip

We endow $\M$ with a vertical positive definite metric, that is a
space--like scalar product
\begin{eqnarray*}
\begin{array}{rcccl}
\Phi : & V(\M) \times_{\M} V(\M) & \to & \Reali &\quad \mbox{s.t.}\\
& ({\bf V}_1 , {\bf V}_2) & \rightsquigarrow & \Phi ({\bf V}_1 , {\bf V}_2) &\quad  \Leftrightarrow\\
& ((t, x^i,V_1^i), (t, x^i,V_2^i)) & \rightsquigarrow & g_{ij} \,
V_1^i \, V_2^j .
\end{array}
\end{eqnarray*}
The vertical metric expresses in a wide sense the embodiment of
the Absolute Space Axiom and usually takes into account the
massive properties of the system. Then the positive definite
matrix $g_{ij}$ is usually called the {\it mass matrix} of the
system. Of course, since $g_{ij} \= \Phi (\frac{\partial}{\partial
x^i}, \frac{\partial}{\partial x^j})$, the expression of $g_{ij}$
depends on the coordinate system.

\begin{Remark} Clearly the vertical metric is not a metric on $\M$ in the
usual sense of Riemannian Geometry, since it acts only on
space--like vectors, and then the mass matrix $g_{ij}$ does not
determine a covariant metric tensor on $\M$. Nevertheless $\Phi$
defines a symmetric double contravariant tensor $\overline{\Phi}$
locally represented by $\overline{\Phi} = g^{ij}
\frac{\partial}{\partial x^i} \otimes \frac{\partial}{\partial
x^j}$, where the matrix $g^{ij}$ is the inverse of the mass matrix
$g_{ij}$. The space--time $\M$ can then be thought of as a
simplified version of the well known concept of Newton--Cartan
space--time (\cite{Cartan1924NC1,Cartan1925NC2,Kunzle1972NC}),
lacking in this case the choice of a connection. \end{Remark}

\begin{Remark} Since an aim of the paper is to illustrate a geometrical
framework for the event--driven algebraic approach to Impulsive
Mechanics, we have to deal only with velocities and impulses
acting on the system, and we do not have to deal with
accelerations and forces (with one exception in Section 3). Then
we do not introduce in details the ``acceleration space--time''
$\JJ$ for the system and its structures and properties. About
these arguments, see for example \cite{MassPaga1991,MassPaga1997}
and the references therein. For the same reason, we do not need
the concept of time--derivative, and then we do not need to
introduce a connection on $\M$.
\end{Remark}

A global frame of reference for $\M$ consists of a
(differentiable) rule giving a one to one correspondence between
each pair of fibres. Each of the following three geometric objects
determines a rule:

\begin{itemize}
\item[-] a global fibred diffeomorphism $\Psi: \M \to \Reali
\times \Q$ (over the map $t:\Euclideo \to \Reali$);

\item[-] a complete time--like vector field ${\bf h}: \M \to \J$;

\item[-] a congruence of world lines for $\M$, i.e. a
$n$--parameter family of global sections
$\sigma_{\xi_1\ldots\xi_n}: \Euclideo \to \M$ such that for every
$m\in \M$ there exists a unique $n$-tuple
$(\overline{\xi}_1\ldots\overline{\xi}_n)$ with $m \, = \,
\sigma_{\overline{\xi}_1\ldots\overline{\xi}_n}(\pi_t(m))$.
\end{itemize}

A congruence of world lines determines a one to one correspondence
between the fibers by setting
\begin{equation*}
\begin{array}{rcccl}
\Psi_{t_0 t_1} : & \{\pi_t^{-1}(t_0)\} & \to & \{\pi_t^{-1}(t_1)\} &\quad \mbox{s.t.}\\
& m_0 = \sigma_{\overline{\xi}_1\ldots\overline{\xi}_n}(t_0) &
\rightsquigarrow &
\sigma_{\overline{\xi}_1\ldots\overline{\xi}_n}(t_1)
\end{array}
\end{equation*}

A diffeomorphism $\Psi: \M \to \Reali \times \Q$ determines a
congruence of world lines  for $\M$ given by the set of
counter--images $\Psi^{-1}(q)$ of the points $q\in \Q$.
Analogously, the integral lines of the vector field ${\bf h}$
determines a congruence of world lines  for $\M$.

It is clearly possible to give a notion of local frame of
reference, renouncing the global definition of the diffeomorphism
$\Psi$ or of the world lines $\sigma$, or renouncing the
completeness of the vector field ${\bf h}$. All the following
arguments still hold in a local sense.

\begin{Remark} The naive choice of $\Euclideo \times \Q$  (or, worst,
$\Reali \times \Q$ or even $\Reali \times \Reali^n$) as
space--time bundle for the system implies an intrinsic choice of a
frame of reference, since the admissible coordinates for
$\Euclideo \times \Q$ have the form $(t, x^1, \ldots, x^n)$ with
transformation group
\begin{eqnarray}\label{VarCoordExQ}
\left\{
\begin{array}{lcl}
\bar{t} &\=& t + const \\
\bar{x}^i & \= & \bar{x}^i(x^1, \ldots, x^n)
\end{array}
\right.
\end{eqnarray}
respecting the natural structures of $\Euclideo \times \Q$ (the
$1$--form $dt$ and the product structure. Then the congruence of
world lines $\sigma_{x^1\ldots x^n}: \Euclideo \to \Euclideo
\times \Q$ such that $\sigma_{x^1\ldots x^n}(t) = (t,x^1, \ldots,
x^n)$ determines a frame of reference for $\Euclideo \times \Q$.

Analogously, the integral lines of the time--like vector field
$\frac{\partial}{\partial t} \in T(\Euclideo\times\Q)$, that is
invariant with respect to the group (\ref{VarCoordExQ}), determine
the (same) frame of reference for $\Euclideo \times \Q$.

This is the reason why the use of naive configuration space--times
like $\Euclideo \times \Q, \Reali \times \Q$ or $\Reali \times
\Reali^n$ are not fit for a frame independent description of the
behavior of the system, introducing and often (possibly
unwittingly) using this intrinsically defined frame of reference.

The unsuitability of the product bundles $\Euclideo \times \Q,
\Reali \times \Q$ and $\Reali \times \Reali^n$ is devastating in
the study of Impulsive Mechanics: in fact velocity, that is the
basic concept in Impulsive Mechanics, is by its very nature a
physical quantity referred to and depending on the knowledge of a
frame of reference. On the other hand, every admissible physical
theory must satisfy the invariance of the formulation with respect
to the frame of reference. Standard requirements of Impulsive
Mechanics, such as for example the preservation of kinetic energy
in an impulsive phenomenon, could be meaningless if the assignment
of the frame of reference where they are formulated is lacking
and, on the other side, it could be not frame invariant if
assigned in a (possibly unwittingly) fixed frame of reference. We
will go back on this argument with explanatory examples in Section
3. \end{Remark}

Later on, unless otherwise specified, a frame of reference will be
considered as a time--like vector field ${\bf h}$, that is a
vector field having the local expression
\begin{equation*}
{\bf h} \=  \dfrac{\partial}{\partial t} + H^1 (t, x^i)
\dfrac{\partial}{\partial x^1}+ \ldots + H^n (t, x^i)
\dfrac{\partial}{\partial x^n} \, .
\end{equation*}
and the set of frames of reference will be denoted with $\H$.

The frame ${\bf h}$, viewed as a section of $\J$, determines a
diffeomorphism, usually called {\it relativization}, $\Delta_{\bf
h} : \J \to V(\M)$ of the affine bundle $\pi:\J \to \M$ with the
vector bundle $\pi:V(\M) \to \M$ such that
\begin{eqnarray*}
\Delta_{\bf h} : \J  \to  V(\M) \quad \mbox{s.t.} \quad
 \Delta_{\bf h}({\bf p}) =  {\bf p} \, - \, {\bf h}(\pi({\bf p})) .
\end{eqnarray*}
The space--like vector $ \vel_{\bf h}({\bf p}) = \Delta_{\bf
h}({\bf p}) = {\bf p} \, - \, {\bf h}(\pi({\bf p}))$ is the {\it
relative velocity} of the absolute velocity ${\bf p}$ {\it with
respect to the frame ${\bf h}$}. The function
\begin{eqnarray*}
\K_{\bf h}: \J  \to  \Reali \quad \mbox{s.t.} \quad \K_{\bf
h}({\bf p})  \= \dfrac12 \Phi \left(\vel_{\bf h}({\bf
p}),\vel_{\bf h}({\bf p})\right)
\end{eqnarray*}
is the {\it kinetic energy}  of the system {\it with respect to
the frame ${\bf h}$}.

\begin{Remark} Once again since one of the aims of the paper is to
illustrate a geometrical framework for Impulsive Mechanics, we are
not interested in the time--derivative $\dfrac{d}{dt}$ determined
by a frame of reference ${\bf h}$ or in the possible connections
${}^{\bf h}\nabla$ determined by ${\bf h}$. About these arguments,
see for example \cite{Pasquero2004ref} and the references therein.
\end{Remark}

\subsection{Impulsive problem for free systems}

The impulsive problem for free systems is easily framed in the
geometric context described above. In fact, roughly speaking, it
consists in the determination of the so--called {\it right
velocity} of the system once a {\it left velocity} of the system
and an impulse acting on the system are known. The intrinsic
character of the impulsive phenomenon, that is independent of the
presence of a frame of reference, implies that the left and
right--velocities are elements ${\bf p}_L, {\bf p}_R$ of $\J$. The
impulse is an element ${\bf I} \in V(\M)$ and the mechanical law
governing the phenomenon is simply the action (\ref{ActionVJ})
\begin{eqnarray}\label{ImpNewEq}
\begin{array}{cccl}
 \J \times V(\M) & \to & \J &\quad \mbox{s.t.}\\
({\bf p}_L , {\bf I}) & \rightsquigarrow & {\bf p}_R \= {\bf p}_L + {\bf I} . \\
\end{array}
\end{eqnarray}
Each frame of reference ${\bf h}$ determines a relativization of
this action so that $({\bf p}_R - {\bf h}) \= ({\bf p}_L - {\bf
h})+ {\bf I}$ or, that is the same, $(\vel_{\bf h})_R - (\vel_{\bf
h})_L \= {\bf I}$ that closely resembles the usual formulation of
the Impulsive Newton Law (INL) and that also exhibits the
independence of the jump of velocities by the frame of reference.

Moreover, the INL in the form (\ref{ImpNewEq}) has the correct
causal structure as equation of the impulsive motion of the free
system. In fact the impulse, in this case called active impulse
${\bf I}_{act}$, can be thought of as a fibred map
\begin{eqnarray}
{\bf I}_{act} :  \J  \to & V(\M) \quad \mbox{s.t.}\quad {\bf p}
\rightsquigarrow  {\bf I}_{act}({\bf p})
\end{eqnarray}
and then the INL assumes the causally correct form
\begin{eqnarray}\label{freeINL}
{\bf p}_R \= {\bf p}_L + {\bf I}_{act}({\bf p}_L).
\end{eqnarray}

\smallskip

Obviously, a free system can have an impulsive behavior only if an
active impulse acts on the system.

\begin{Remark} Eq. (\ref{freeINL}) is unassailable from the theoretical point of
view. However, the principle of determinism of Classical Mechanics
and the causal structure of (\ref{freeINL}) imply that, in order
to make it effective as equation of motion of the system, the
active impulse must be an {\it a priori} assignment.
Unfortunately, giving a physical meaning to the concept of active
impulse could be a hard task. In fact, note that an active impulse
cannot be globally defined on the whole $\J$ (since otherwise
(\ref{freeINL}) should be endlessly applied in any instant). We
could introduce a submanifold $\A$ of $\J$ with the property that,
if ${\bf p} \in \A$ then ${\bf p} + {\bf I}_{act}({\bf p}) \notin
\A$ (since otherwise the INL should be applied repentantly) but,
although this requirement is of course mathematically correct, the
manifold $\A$ has a difficult physical interpretation (without
making use of the concept of constraint) and meaningful physical
systems subject to active impulses rarely appear in literature (if
not as subsystems of constrained systems).

We do not delve into a detailed analysis of the effective
applicability of the INL (\ref{freeINL}) (see
\cite{Pasquero2018AAPP}). However we anticipate that the
applicability problem does not appear when the impulsive behavior
of the system is due to the presence of constraints.
\end{Remark}

\section{Geometry of constrained systems}

A constraint acting on a system is, in the widest sense, any
limitation imposed on the possible motions of the system. The
extreme generality of this statement is usually clarified by
distinguishing different types of constraints: for instance,
positional or kinetic, bilateral or unilateral, permanent or
instantaneous, isotropic or not, ideal or not, with or without
friction, breakable or unbreakable. Only some of these
distinctions are of geometric nature, and they are described in
this section. Moreover, of course, only some combinations of these
types of constraints have a clear physical meaning.

In order to make effectively applicable the general definition and
the consequent classification, moreover abiding by the physical
meaning of the constraint, we introduce additional geometric
structures on the framework set up for free systems (see
\cite{Pasquero2005uni,Pasquero2006,Pasquero2008,Pasquero2012}).

\begin{Remark} The causal structure of the Newton's Second Law
implies that constraints on a system can be assigned only as
limitations on its admissible positions and/or velocities.
Limitations assigned on the admissible accelerations of the system
are not in general considered as constraints, but are considered
as controls (and they are the central argument of Control Theory).
Once again then we have no need to introduce the ``acceleration
space--time'' $\JJ$.
\end{Remark}

\subsection{Geometry of positional constraints}
A positional constraint consists, roughly speaking, is a
constraint on the admissible space--time configuration of the
system. It can be modelled with a fibred subbundle $i:\S \to \M$
that at this stage we suppose of constant dimension $r+1$ with
$0<r<n$, and without boundary. The bundle $\S$ determines the
following additional geometric objects and structures relative to
the system:

\begin{itemize}
\item[0)] the local description of $\S$, that can be given in
parametric form $x^i = x^i(t,q^1,\ldots,q^r), \ i=1,\ldots,n$, or
in cartesian form $f_{\rho}(t,x^1,\ldots,x^n)=0,
\rho=1,\ldots,n-r$.

\item[1)] The affine subbundle $i_*:\JS \to \J$  of the absolute
velocities that are tangent to $\S$.

\item[2)] The vector subbundle $i_*:V(\S) \to V(\M)$ of the
vertical vectors that are tangent to the fibers of $\S$

\item[3)] The pull--back bundle $\pi: (i_*)^*(\J) \to \S$, that is
the {\it velocity space--time} of the system when the system is
{\it in contact} with the constraint $\S$, and the pull--back
bundle $\pi: (i_*)^*(V(\M)) \to \S$, that is the bundle of the
space vectors of the system when the system is {\it in contact}
with the constraint $\S$. Both the pull--back bundles
$(i_*)^*(\J)$ and $(i_*)^*(V(\M))$ are locally described by
admissible coordinates
$(t,q^1,\ldots,q^r,\dot{x}^1,\ldots,\dot{x}^n)$.

\item[4)] Thanks to the presence of the vertical metric $\Phi$,
the splitting $(i_*)^*(V(\M)) \= i_*(V(\S)) \oplus
i_*(V^{\perp}(\S))$ with its associated projection operators $\PV,
\PVO$.

\item[5)] The splitting $(i_*)^*(\J) \= i_*(\JS) \oplus
i_*(V^{\perp}(\S))$, where in this case the direct sum $\oplus$
reflects the action (\ref{ActionVJ}),  and its associated
projection operators $\PJ, \PJO$.

\item[6)] The subclass $\H_{\S}$ of the frames of reference ${\bf
h}_{\S}$ of $\M$ such that $\left({\bf
h}_{\S}\right)_{\rfloor_{\S}}$ is
 tangent to $\S$. The
elements of $\H_{\S}$ are called the (possible) {\it rest frames}
of $\S$. If $\S$ is assigned in cartesian form
$f_{\rho}(t,x^1,\ldots,x^n)=0$, they are characterized by the
conditions ${\bf h}_{\S}(f_{\rho})=0 \, \forall
\rho=1,\ldots,n-r$.

\end{itemize}

\begin{Example} Let $\M$ be the configuration space--time of a
massive particle freely moving in a $3$--dimensional euclidean
space, so that $\M = \Euclideo \times \Euclideo^3$. If $\M$ is
described by cartesian coordinates $(t,x,y,z)$, let $\S_1$ be the
subbundle described by the immersion $i:\S_1 \to \M$ such that
$(t,q^1,q^2) \rightsquigarrow (t,x=q^1,y=q^2,0)$ or alternatively
by the cartesian representation $z=0$. \end{Example}

\begin{Example} Let $\M$ be as above and let $\S_2$ be the subbundle
described by the immersion $i:\S_2 \to \M$ such that $(t,q)
\rightsquigarrow (t,x=q,0,0)$ or alternatively by the cartesian
representation $y=z=0$. \end{Example}

\subsubsection{Bilateral positional constraint}
If we require that all the possible motions $\gamma:\Euclideo \to
\M$ of the system obey the condition $\gamma:\Euclideo \to \S
\subset \M$, then all the admissible configurations for the system
belong to $\S$ (and all the possible absolute velocities of the
system belong to $\JS$ and are then tangent to $\S$). In this
case, $\S$ represents a bilateral positional permanent constraint.
Such a kind of constraint, depending on its action on the system,
could be absorbed in the construction itself of the space--time
bundle $\M$. We will briefly go back on this kind of constraints
along the paper, but we are not in general interested in these
constraints, except when they have an impulsive character.

\medskip

\subsubsection{Unilateral positional constraint}
The pull--back bundle $(i_*)^*(\J)$ is formed by the time--like
vectors ${\bf p} \in \J$ such that $\pi({\bf p}) \in i(\S)$, but
${\bf p}$ is not necessarily tangent to $\S$: then $(i_*)^*(\J)$
is the natural geometric framework fit to analyze the behavior of
$\S$ viewed as a unilateral positional constraint.

\begin{Definition}\label{VincoloUnilatero}
A positional constraint $\S$ is called unilateral in a point $s
\in \S$ if two sets $\L_s(\J) \subset (i_*)_{s}^*(\J)$ and
$\R_s(\J)\subset (i_*)_{s}^*(\J)$  are assigned so that the space
$(i_*)_{s}^*(\J)$ of the time--like vector of $\M$ applied in $s$
can be written as the disjoint union
\begin{eqnarray}\label{DecomposizioneLR}
(i_*)_{s}^*(\J) \= \L_s(\J) \cup (i_*)_s(\JS) \cup \R_s(\J)
\end{eqnarray}

The constraint $\S$ is called unilateral if it is unilateral in
every point $s \in \S$.
\end{Definition}

The set $\L_s(\J)\cup (i_*)_s(\JS)$ is the set of the {\it
admissible left velocities} for the system in contact with $\S$ in
the point $s\in \S$, while the set $\R_s(\J)\cup (i_*)_s(\JS)$ is
the set of the {\it admissible right velocities} for the system in
contact with $\S$ in the point $s\in \S$. We define $\L_{\S}(\J) =
\bigcup_{s\in\S} \L_s(\J), \, \R_{\S}(\J) = \bigcup_{s\in\S}
\R_s(\J)$. The set $\L_{\S}(\J)$ is also called the set of {\it
incoming} or {\it entrance} velocities, while the set
$\R_{\S}(\J)$ is that of {\it outgoing} or {\it exit} velocities.

\medskip

The constraint $\S_1$ of the previous Ex.1 has a natural
unilateral nature, while the constraint $\S_2$ of Ex.2 does not.
However, to justify in detail this statement we need to delve into
the geometric structure of $(i_*)^*(\J)$.

\medskip

An important result about the geometric structures determined by
the subbundle $\S$ and the contact bundle $(i_*)^*(\J)$ is the
following (see \cite{Pasquero2005uni}):
\begin{Theorem}\label{VelOrt}
Given a frame of reference ${\bf h} \in \H_{\S}$, the diagrams
\begin{eqnarray*}
\hskip-7truemm
\begin{array}{lcl} \xymatrix @R=15pt @C=10pt{
(i_*)^*(\J) \ar |(0.5){\PJO} [ddd] \ar |(0.5){\Delta_{\bf
h}}[rrr]&&& (i_*)^*(V(\M))  \ar |(0.5){\PVO}
[ddd]\\
& \\& \\
i_*(V^{\perp}(\S))\,\ar@{=}[rrr]&&& i_*(V^{\perp}(\S)) } & \quad &
\xymatrix @R=15pt @C=10pt{ (i_*)^*(\J) \ar |(0.5){\PJ} [ddd] \ar
|(0.5){\Delta_{\bf h}}[rrr]&&&(i_*)^*(V(\M))  \ar |(0.5){\PV}
[ddd]\\
& \\& \\
i_*(\JS)\,\ar |(0.5){\Delta_{\bf h}}[rrr] &&&i_*(V(\S)) }
\end{array}
\end{eqnarray*}
commutes. The diagrams does not commute if ${\bf h} \notin
\H_{\S}$.
\end{Theorem}

The theorem clarifies the frame invariance properties of the
orthogonal and parallel components of an absolute velocity with
respect to a positional constraint.

Given an absolute velocity ${\bf p} \in (i_*)^*(\J)$ and a generic
frame ${\bf h} \in \H$ we can construct the two space--like
vectors $\vel_1^{\perp}({\bf p}) \= \PJO({\bf p})$, and
$\vel_2^{\perp}({\bf p}) \=  \PVO({\bf p} - {\bf h})$. Both are
elements of $(i_*)(V^{\perp}(\S))$: the first is manifestly
independent of the frame ${\bf h}$ while the second in general
depends on ${\bf h}$, so that in general $\vel_1^{\perp}({\bf p})
\neq \vel_2^{\perp}({\bf p})$.  The theorem states that the two
vectors coincide if and only if ${\bf h}$ is a rest frame for
$\S$. Then the best definition of orthogonal component
$\velort({\bf p})$ of the absolute velocity ${\bf p}$ with respect
to $\S$ can be given only for frames in the class $\H_{\S}$, and
it is $\velort({\bf p}) \= \PJO({\bf p}) \=  \PVO({\bf p} - {\bf
h}_{\S})$.

In a similar way, given ${\bf p} \in (i_*)^*(\J)$ and ${\bf h} \in
\H$ we can construct the two space--like vectors $\vel_1^{\|}({\bf
p}) \= \PJ({\bf p}) - {\bf h}$ and $\vel_2^{\|}({\bf p}) \=
\PV({\bf p} - {\bf h})$: the second is an element of $i_*(V(\S))$
for every ${\bf h} \in \H$, while the first in general is an
element of $(i_*)^*(V(\M))$ and $\vel_1^{\|}({\bf p}) \in
i_*(V(\S))$ if and only if ${\bf h} \in \H_{\S}$. Then once again
in general $\vel_1^{\|}({\bf p}) \neq \vel_2^{\|}({\bf p})$. The
theorem states that the two vectors coincide if and only if ${\bf
h}$ is a rest frame for $\S$. Then once again the best definition
of parallel component $\vel^{\|}_{\bf h}({\bf p})$ of the absolute
velocity ${\bf p}$ with respect to $\S$ can be given only for
frames in the class $\H_{\S}$, and it is $\vel^{\|}_{\bf h}({\bf
p}) \= \PJ({\bf p}) - {\bf h}_{\S} \= \PV({\bf p} - {\bf
h}_{\S})$.

\begin{Remark} Note that, even if we restrict the choice of frames in the
class $\H_{\S}$, the parallel component depends on the frame ${\bf
h}$. This will play a crucial role in the next section especially
regarding the concept of friction. \end{Remark}

\begin{Example} If a massive particle moving in the $3$--dimensional
euclidean space impacts with a plane, the orthogonal component of
the impact velocity is univocally determined by the geometry of
the system (the plane and its class of rest frames), while the
tangent component of the velocity is not univocally determined. In
fact, naively speaking, the tangent component of the velocity
changes if, in the contact point, the plane is at rest or if it is
formed, for example, by a conveyor belt.
\end{Example}

It is a straightforward matter that, if $\S \subset \M$ is of
codimension $1$, then $(i_*)^*(V^{\perp}(\S))$ has dimension $1$
and we can choose a (possibly but not necessarily unit) vector
${\bf U}^{\perp}$ such that $(i_*)^*(V^{\perp}(\S)) = Lin({\bf
U}^{\perp})$. Then, for every ${\bf p} \in (i_*)^*(\J)$, we can
evaluate the sign of $\Phi( \velort({\bf p}), {\bf U}^{\perp})$:
recalling that, of course, $\Phi( \velort({\bf p}), {\bf
U}^{\perp}) = 0$ implies $\velort({\bf p}) = 0$ and so ${\bf p}
\in i_*(\JS)$, we can define, for example,
\begin{eqnarray}\label{CriterioLR}
\begin{array}{l}
{\bf p} \in \L_{\S}(\J) \Leftrightarrow \Phi( \velort({\bf p}),
{\bf U}^{\perp}) < 0 \\
{\bf p} \in \R_{\S}(\J) \Leftrightarrow \Phi( \velort({\bf p}),
{\bf U}^{\perp}) > 0
\end{array}
\end{eqnarray}

\begin{Example} Going back to the previous examples 1 and 2, the
constraint $\S_1$ is naturally unilateral because it is of
codimension $1$. We can set ${\bf U}^{\perp} =
\frac{1}{\sqrt{m}}\frac{\partial}{\partial z}$ (or simply ${\bf
U}^{\perp} =\frac{\partial}{\partial z}$) so that
$(i_*)^*(V^{\perp}(\S_1)) = Lin({{\bf U}^{\perp}}) =
Lin(\frac{\partial}{\partial z})$ and then the rule
(\ref{CriterioLR}) determines the admissible left and right
velocities for the particle when in contact with $\S_1$.

Differently, the constraint $\S_2$ is not naturally unilateral.
Since, roughly speaking, the constraint consists in a
$1$--dimensional line in a $3$--dimensional space, every
non--tangent velocity of the particle in contact with $\S_2$ can
be either an entrance or an exit velocity for the particle. Of
course, we can choose, although in an arbitrary way, a splitting
(\ref{DecomposizioneLR}). \end{Example}

\begin{Remark} It goes without saying that an effective geometrization of
unilateral constraints allows the construction of an effective
geometric model of positional constraints with boundary. If we
choose ${\bf U}^{\perp}=\frac{\partial}{\partial z} $ and
$\L_{\S_1}(\J), \R_{\S_1}(\J) $ defined as in (\ref{CriterioLR}),
the constraint $\S_1$ of Ex.1 is the geometric model of the
positional constraint with boundary given by the condition
$z\ge0$. However, the constraint $\S_2$ together with an arbitrary
assignment of $\L_{\S_2}(\J)$ and $\R_{\S_2}(\J) $, due to its
codimension greater than $1$, has not a clear counterpart in terms
of positional constraint with boundary. We do not analyze in depth
this argument, that will be simply sketched in the following
examples. \end{Remark}

\subsubsection{Multiple positional constraints}
It is clear that the presence of two ore more bilateral positional
constraints $\S_1, \S_2$ can be modelled with a single bilateral
positional constraint given by their intersection $\S_1 \cap
\S_2$. Then we restrict our attention to multiple unilateral
positional constraints.

Def. \ref{VincoloUnilatero} allows to highlight the difference
between a ``genuine'' constraint of codimension greater than $1$
and the simultaneous action of more than one constraint of
codimension $1$. Let us refer to Ex.1 and Ex.2.

\begin{Example} Let $\M$ be the space--time bundle of Ex.1 and Ex.2,
and, with a slight abuse of notation, let $\S_y, \S_z$ be the
subbundles described respectively  by the immersions $i_y:\S_y \to
\M$ such that $(t,x,z) \rightsquigarrow (t,x,0,z), \, i_z:\S_z \to
\M$ such that $(t,x,y) \rightsquigarrow (t,x,y,0)$ or
alternatively by the respective cartesian representations $\S_y =
\{y=0\}, \S_z=\{z=0\}$. Then of course we have $\S_2 = \S_y \cap
\S_z$, but, although $\S_2$ and $\S_y \cap \S_z$ are the same
subbundle of $\M$, from the mechanical point of view they have
different behaviors. In fact we already saw that the constraint
$\S_2$ is not naturally unilateral since it does not admit a
natural choice of entrance and exit velocities. On the contrary,
the constraint $\S_y \cap \S_z$ allows a natural splitting
(\ref{DecomposizioneLR}): with obvious notation, we  have
$(i_*)^*(V^{\perp}(\S_y)) = Lin(\frac{\partial}{\partial y}),
(i_*)^*(V^{\perp}(\S_z)) = Lin(\frac{\partial}{\partial z})$ and
we can define
\begin{eqnarray}\label{CriterioLRMultiplo}
\begin{array}{l}
{\bf p} \in \L_{\S_y \cap \S_z}(\J) \Leftrightarrow \Phi(
\velort_{\S_y}({\bf p}), \frac{\partial}{\partial y}) < 0 \quad
\mbox{or} \quad \Phi( \velort_{\S_z}({\bf p}),
\frac{\partial}{\partial z}) < 0\\ \\{\bf p} \in \R_{\S_y \cap
\S_z}(\J) \Leftrightarrow \left\{
\begin{array}{l}
\Phi( \velort_{\S_y}({\bf p}), \frac{\partial}{\partial y}) \ge 0 \\ \\
\Phi( \velort_{\S_z}({\bf p}), \frac{\partial}{\partial z}) > 0
\end{array} \right.
\quad \mbox{or} \quad \left\{
\begin{array}{l}
\Phi( \velort_{\S_y}({\bf p}), \frac{\partial}{\partial y}) > 0 \\ \\
\Phi( \velort_{\S_z}({\bf p}), \frac{\partial}{\partial z}) \ge 0
\end{array} \right.
\end{array}
\end{eqnarray}
Then the rule (\ref{CriterioLRMultiplo}) determines the admissible
left and right velocities for the particle when in contact with
$\S_y \cap \S_z$.

The unilateral constraints $\S_y$ and $\S_z$, together with the
condition (\ref{CriterioLRMultiplo}), can be considered the model
of the unilateral constraint with boundary given by
$\left\{\begin{array}{l}y\ge 0 \\ z \ge 0
\end{array}\right.$. Instead, the constraint $\S_2$ does not give positional restrictions
to the particle. \end{Example}

\begin{Definition}
A multiple unilateral positional constraint $\S$ is a regular
intersection of unilateral positional constraints $\S_i,
i=1,\ldots,r \ge 2 $ of codimension $1$. The intersection is
regular if the  vectors $\{{\bf U}^{\perp}_{\S_i}, i=1,\ldots,r\}$
are linearly independent in every point of $\S = \bigcap_{
i=1,\ldots,r} \S_i$.
\end{Definition}

\subsubsection{Anisotropic positional constraints}
The anisotropy of $\S$ consists, roughly speaking, in the
existence of one or more ``special directions'' tangent to $\S$,
in general reflecting peculiar physical properties of $\S$ (such
as the presence of climbing skins, riflings...). However, we
already saw (see Th. \ref{VelOrt}) that the assignment of the
subbundle $\S$ alone does not determine natural tangent directions
on $\S$. Then they must be assigned as an additional structure.

The simplest way to determine a distinguished direction on $\S$
consists in the assignment of a  section ${\bf L}_{Anis}: \S \to
V(\S)$ of the vertical bundle of $\S$ assigning to each point $s
\in \S$ a space--like vector ${\bf L}_{Anis}(s) \in V_s(\S)$
tangent to the fiber of $\S$. Since $V(\S) \subset V(\M)$, in
order to point out the directional and not the vectorial property
of ${\bf L}_{Anis}$, we can require the unitary condition $\|{\bf
L}_{Anis} \|^2 = \Phi({\bf L}_{Anis},{\bf L}_{Anis}) =1$.

\begin{Example} The constraint $\S_1$ of Ex.1 together with the unit
space--like vector field ${\bf L}_{Anis} = \frac{1}{\sqrt{m}}
\frac{\partial}{\partial q^1}$ in an anisotropic constraint with a
field of distinguished directions. \end{Example}

\begin{Remark} A formal assignment of a distinguished direction on $\S$
should involve the $1$--dimensional Grassmanian manifold of $\S$.
If the anisotropy of the constraint depends on structures more
complicated than directions (such as planes), we can assign two or
more unit space--like vector fields generating the structures, or
sections of the multi--dimensional Grassmanian manifolds of $\S$.
\end{Remark}

\subsection{Geometry of kinetic constraints}
Kinetic constraints are, roughly speaking, those that fix
limitations on the admissible velocities of the system without
fixing limitations on its configurations.

\subsubsection{Permanent kinetic constraints}
A permanent kinetic constraint is a kinetic constraint that acts
on the system before, during and after the instant of the
impulsive behavior. A permanent kinetic constraint $\A$ can be
modelled with a fibred subbundle $i:\A \to \J$.

The affine structure of the constraint $\A$ viewed as subbundle of
$\J$ is a usual requirement in order to ensure that the impulsive
problem holds on to be governed by an INL of type (\ref{ActionVJ})
representing the action of a modelling vector bundle on an affine
bundle. In this case in fact, the differences ${\bf A} \= a_1 -
a_2, a_i \in \A$ between elements of $\A$ form a vector subspace
$V(\A)$ of $V(\M)$. However this requirement does not reflect a
mandatory condition of Impulsive Mechanics and non--linear kinetic
constraints are likewise framed in this geometric context.

\begin{Remark} A classical argument about permanent kinetic constraints is
their integrability, that is if $\A = \JS$ for some positional
constraint $\S$. Several results are known about this problem
(see, e.g. \cite{MassPaga1997}), but they are not in the aims of
this paper. Then we restrict our attention (with one exception in
this Section) to ``genuine'' permanent kinetic constraints (that
is, kinetic constraints that are not kinetic formulations of
positional constraints). Moreover, the conditions for which a
kinetic constraint admits an affine description is once again a
classical argument and several results are known about this
problem (see once again \cite{MassPaga1997}), but they are not in
the aims of this paper. \end{Remark}

\begin{Example} The pure rolling conditions for a sphere of radius $R$
constrained to move in contact with a horizontal plane is an
example of permanent kinetic constraint. Describing the
space--time $\M$ with the usual coordinates
$(t,x,y,\psi,\vth,\vph)$ where $(x,y)$ are the coordinates of the
center of the sphere and $(\psi,\vth,\vph)$ are the Euler angles,
the permanent kinetic constraint $\A$ can be given by the linear
(in the dotted coordinates) equations
\begin{eqnarray}\label{vincolo cinetico piano}
\left\{
\begin{array}{l}
\dot{x} - R \dot{\vth}\sin\psi + R \dot{\vph} \sin\vth\cos\psi \= 0 \\
\dot{y} + R \dot{\vth}\cos\psi + R \dot{\vph} \sin\vth\sin\psi \=
0 .
\end{array}
\right.
\end{eqnarray}
\end{Example}

If $\A$ is an affine subbundle of $\J$, the vector subbundle
$i_*:V(\A) \to V(\M)$ determines, thanks to the presence of the
vertical metric $\Phi$,  the splitting $V(\M) \= V(\A) \oplus
V^{\perp}(\A)$ with its associated projection operators, and the
splitting $\J \= \A \oplus V^{\perp}(\A)$ with its associated
projection operators. In particular, given ${\bf p} \in \J$, we
have that ${\bf p} = \P_{\A}^{\|}({\bf p}) + \velortA({\bf p})$.
It is a straightforward corollary of Th. \ref{VelOrt} that the
space--like vector $\velortA({\bf p})$ is invariant with respect
to the subclass $\H_{\A}$ of frames of $\H$ having image in $\A$
(that is, naively speaking, the class of the rest frames of $\A$).

\begin{Remark} It is possible to assign an impulsive problem for a
mechanical system by assigning a kinetic constraint $\A \subset
\J$ such that, for a fixed instant $t_0$ or for the points of an
assigned subset $\N \subset \M$, $\A \= \R_{\N}(\J)$ is the set of
admissible right velocities (while $\L_{(\M \setminus \N)}(\J) \=
\J$ itself). This is the well known case of the so--called {\it
inert constraints} (see e.g.
\cite{Ibortetal2000,Pasquero2005Carnot}).
\end{Remark}

\subsubsection{Instantaneous kinetic constraints}
An instantaneous kinetic constraint is a kinetic constraint that
acts on the system only during the instant of the impulsive
behavior, usually the instant of collision of the system with a
positional constraint $\S$. An instantaneous kinetic constraint
$\B$ is then modelled with a fibred subbundle $i:\B \to \JS$.

Once again the affine structure of the constraint $\B$ viewed as
subbundle of $\JS$ is a usual requirement for $\B$, so that the
differences ${\bf B} \= b_1 - b_2, b_i \in \B$ between elements of
$\B$ form a vector subspace $V(\B)$ of $V(\S)$. In this case, the
INL (\ref{ActionVJ}) gives once again the equation of motion of
the system.

\begin{Example} The pure rolling conditions in the contact point for a
sphere of radius $R$ moving in a $3$--dimensional halfspace and
impacting with a horizontal plane is an example of instantaneous
kinetic constraint. Describing the space--time $\M$ with the usual
coordinates $(t,x,y,z,\psi,\vth,\vph)$ where $(x,y,z)$ are the
coordinates of the center of the sphere and $(\psi,\vth,\vph)$ are
the Euler angles, the positional constraint $\S$ is given by the
condition $z-R=0$, and the instantaneous kinetic constraint $\B$
is given by the equations (\ref{vincolo cinetico piano}).
\end{Example}

If $\B$ is an affine subbundle of $\JS$, the vector subbundle
$i_*:V(\B) \to V(\S)$ determines the splitting $V(\S) \= V(\B)
\oplus V^{\perp}(\B)$ with its associated projection operators,
and the splitting $\JS \= \B \oplus V^{\perp}(\B)$ with its
associated projection operators. Taking into account the immersion
$i:\S \to \M$, we obtain the splittings $V(\M) \= V(\B) \oplus
V^{\perp}(\B) \oplus V^{\perp}(\S)$ and $\J \= \B \oplus
V^{\perp}(\B) \oplus V^{\perp}(\S)$. In this case, given ${\bf p}
\in  (i_*)^*(\J)$, we have that ${\bf p} = \P_{\B}^{\|}({\bf p}) +
\velortB({\bf p}) + \velortS({\bf p})$, and we already discussed
the invariant properties of $\velortS({\bf p})$ with respect to
the class $\H_{\S}$ of rest frames of $\S$. It is a
straightforward corollary of Th. \ref{VelOrt} that the space--like
vector $\velortB({\bf p})$ has the same invariant properties with
respect to the subclass $\H_{\B}$ of frames of $\H_{\S}$ having
image in $\B$ (that is, naively speaking, the class of the rest
frames of $\B$).

\begin{Remark} If the instantaneous kinetic constraint $\B$ has
codimension $1$ in the affine bundle $\JS$, then the vector bundle
$V^{\perp}(\B)$ has dimension $1$, and so there exists a
space--like unit vector field ${\bf U}_{\B}^{\perp}$ such that
$V^{\perp}(\B)= Lin\{ {\bf U}_{\B}^{\perp} \}$. Then the
assignment of $\B$ implies the assignment of a direction of
anisotropy of $\S$. \end{Remark}

\subsubsection{Further remarks on kinetic constraints}

Kinetic constraints are so naturally embodied in the geometric
setup that their definition can be easily extended to unilateral
cases: for instance, the kinetic conditions (\ref{vincolo cinetico
piano}) expressing the pure rolling of a sphere on a horizontal
plane can be modified in the form
\begin{eqnarray}\label{vincolo cinetico unilatero2}
\left\{
\begin{array}{l}
\dot{x} - R \dot{\vth}\sin\psi + R \dot{\vph} \sin\vth\cos\psi \ge 0 \\
\dot{y} + R \dot{\vth}\cos\psi + R \dot{\vph} \sin\vth\sin\psi \ge
0 \, .
\end{array}
\right.
\end{eqnarray}
Of course being the conditions (\ref{vincolo cinetico unilatero2})
mathematically correct, the physical meaning of such  a constraint
is hard to conceive. However, the following mechanical system
gives an example of permanent kinetic unilateral constraint.

\begin{Example} A disk moves with its boundary in contact with
a horizontal rough plane and with the axis of the disk kept in
horizontal position. The system can be described by coordinates
$(t,x,y,\vth,\vph)$ where $(x,y)$ are the coordinates of the
center of the disk, $\vth$ is the orientation of the vertical
plane of the disk with respect to a fixed vertical plane and
$\vph$ is the orientation of the disk with respect to a horizontal
plane containing its axis. The disk is subject to a coaster brake,
so that the condition $\dot{\vph}\ge 0$ holds. If moreover the
disk is subject to the pure rolling kinetic constraint, $\A$ is
expressed by the conditions
\begin{eqnarray*} \left\{
\begin{array}{lcl}
\dot{x} + R \dot{\vph} \cos \vth & \= & 0
\\
\dot{y} + R \dot{\vph} \sin \vth & = & 0 \\
\dot{\vph} & \ge & 0
\end{array}
\right.
\end{eqnarray*}
\end{Example}

It is however important to note that, at odds with the naturalness
of the formal definition of kinetic constraints, the effective
assignment of the kinetic restrictions exerted on a mechanical
system could require several details and clarifications. This
happens even for the standard example of pure rolling conditions.
In fact the pure rolling conditions consist in the requirement
that ``the points of the system and of the constraint where the
contact happens'' have the same velocity (see, e.g.
\cite{LeviCivita}). Nevertheless we know that the concept of
velocity can be the absolute one, that is ${\bf p} \in \J$, or the
relative one (with respect to an assigned frame ${\bf h}$), that
is ${\bf V} = {\bf p} - {\bf h} \in V(\M)$.

It is clear that both ``the velocities of the contact point of the
system and of the contact point of the constraint'' cannot be
elements of $\J$. In fact the geometry of the space--time bundle
$\M$ could be such that ``the point of the constraint where the
contact happens'' does not even exists. This is the case of Ex.7,
where the pure rolling conditions are referred to a phantom
constraint $z-R=0$ (the one of Ex.8) that however is not
representable using the geometry of $\M$.

Moreover, note that even if a permanent bilateral positional
constraint $z-R=0$ is introduced in Ex.7, it is not possible to
describe potential ``inner'' motions of the plane $z-R=0$ by using
only elements of $\J$: a steady horizontal plane and a
horizontally moving conveyor belt are impossible to discern
without the introduction of frames of reference. Therefore the
explicit formulation of pure rolling conditions can be made only
by assigning a frame of reference where the velocities can be
evaluated. It should be then proved that the pure rolling
conditions are independent of the frame chosen to describe them.
This proof requires a detailed description of Relative Mechanics
expressed using the geometric context of fibred bundles and the
concept of frame of reference viewed as a vector field. About
these arguments see for example \cite{Pasquero2004ref}.

\subsection{Miscellaneous}

The simultaneous presence of constraints of the same kind acting
on the system was already taken into account by introducing
multiple constraints in the case of unilateral positional
constraints and by the definition itself of (permanent or
instantaneous) kinetic constraints. However, an impulsive system
can be simultaneously subject to different kinds of constraints.

\begin{Example} In a classical ``billiard situation'', a sphere
rolling on a horizontal plane impacts with a vertical wall, with a
pure rolling condition of the sphere in the contact point of the
vertical wall during the impact. This is a very natural example of
system simultaneously subject to three different kinds of
constraints: a unilateral positional constraint $\S$ (the vertical
wall), a permanent kinetic constraint $\A$ (the pure rolling
condition on the horizontal floor) and an instantaneous kinetic
constraint $\B$ (the pure rolling condition in the contact point
of the sphere with the vertical wall). \end{Example}

The simultaneous presence of unilateral positional constraints
$\S$ and permanent kinetic constraints $\A$ provides the geometric
context with all the structures determined by each constraint and
with the structures that can be constructed with them.

\begin{Example} A disk of radius $R$ moves in a vertical
plane rolling on a horizontal line and impacts with a vertical
wall. The space--time bundle $\M$ can be locally described by
admissible coordinates $(t,x,\vartheta)$ where $x$ is the
cartesian abscissa of the center of the disk in the plane of the
motion and $\vartheta$ is the orientation of the disk. The
unilateral constraint $\S$ is given by the condition $x\=
constant$ while the kinetic constraint $\A$ is given by the
(integrable) condition $\dot{x} + R \dot{\vth} \= 0$.  We have
then, for example,
\begin{equation*}
\begin{array}{l}
\bullet \,\, {\bf h} \, \in \, \H_{\S} \,\,\, {\rm if \, and \,
only \, if} \,\,\, {\bf h} \= \dfrac{\partial}{\partial t}  +
\dot{\vartheta}
\dfrac{\partial}{\partial \vth} \\ \\
\bullet \,\, {\bf h} \, \in \H_{\S}\cap\H_{\A} \,\,\, {\rm if \,
and \, only \, if} \,\,\, {\bf h} \= \dfrac{\partial}{\partial t}
\\ \\%
\bullet \,\, \L_{\S}(\J) \= \left\{ \dfrac{\partial}{\partial t} +
\dot{x} \dfrac{\partial}{\partial x} + \dot{\vartheta}
\dfrac{\partial}{\partial \vth} \, \Big| \, \dot{x} > 0 \right\} \\ \\ %
\bullet \,\, \L_{\S}(\A) \= \left\{ \dfrac{\partial}{\partial t} -
R \dot{\vartheta} \dfrac{\partial}{\partial x} + \dot{\vartheta}
\dfrac{\partial}{\partial \vth} \, \Big| \, \dot{\vth} <0 \right\}
\= \L_{\S}(\J) \cap \A
\end{array}
\end{equation*}
If ${\bf p}\= \frac{\partial}{\partial t} - R \dot{\vartheta}
\frac{\partial}{\partial x} + \dot{\vartheta}
\frac{\partial}{\partial \vth} \, \in \, \A$ is an admissible
velocity of the disk, we have also
\begin{equation*}
\begin{array}{l}
\bullet \,\,  \velortS({\bf p}) \= - R \dot{\vartheta}
\dfrac{\partial}{\partial x} \\ \\ %
\bullet \,\, \velort_{\JS\cap\A}({\bf p}) \= - R \dot{\vartheta}
\dfrac{\partial}{\partial x} + \dot{\vartheta}
\dfrac{\partial}{\partial \vth} %
\end{array}
\end{equation*}
\end{Example}

\subsection{The global diagram}

The whole geometric construction fitting to frame the impulsive
mechanical problem for a constrained system is then the following:
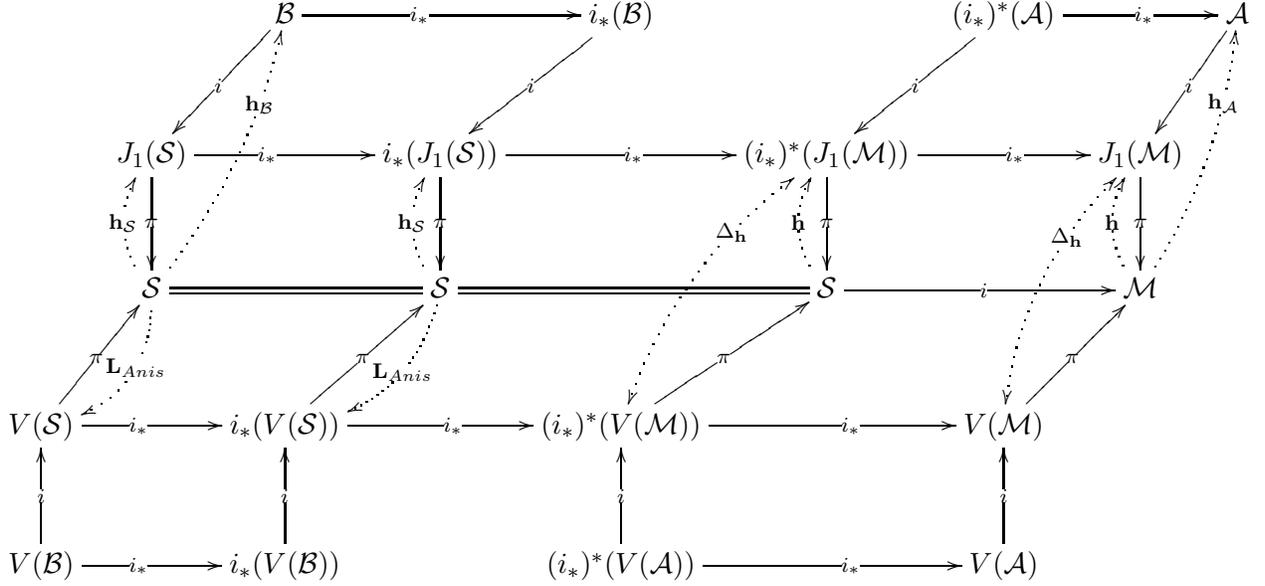
\begin{equation*}
\hskip-2truecm
\xymatrix @R=15pt @C=10pt{%
&&\B \ar |(0.5){i} [ddl] \ar |(0.4){i_*}[rr] &&i_*(\B) \ar
|(0.5){i} [ddl] & & (i_*)^*(\A)\ar |(0.5){i} [ddl]
\ar |(0.6){i_*}[rr]&& \A \ar |(0.5){i} [ddl]\\ \\%
&\JS \ar |(0.5){\pi} [dd] \ar |(0.4){i_*}[rr]&&\iJS \ar
|(0.5){\pi} [dd]\ar |(0.5){i_*}[rr]&&\IJS \ar |(0.5){\pi} [dd]
\ar |(0.6){i_*}[rr]&& \J \ar |(0.5){\pi} [dd]& \\ \\%
&\S\ar@{=}[rr] \ar@/^-.6pc/@{..>} |(.7){{\bf h}_{\B}}[uuuur]
\ar@/^1.2pc/@{..>} |(.4){{\bf L}_{Anis}}[ddl] \ar@/^.9pc/@{..>}
|(.5){{\bf h}_{\S}}[uu] && \S\ar@{=}[rr] \ar@/^.9pc/@{..>}
|(.5){{\bf h}_{\S}}[uu] \ar@/^1.2pc/@{..>} |(.4){{\bf
L}_{Anis}}[ddl] &&\S\ar |{i} [rr] \ar@/^.9pc/@{..>} |(.5){\bf
h}[uu] && \M \ar@/^.9pc/@{..>} |(.5){\bf h}[uu]
\ar@/^-.6pc/@{..>} |(.7){{\bf h}_{\A}}[uuuur]& & \\ \\%
V(\S)\ar |(0.5){\pi} [uur] \ar |(0.4){i_*}[rr] &&\iV \ar
|(0.5){\pi} [uur] \ar |(0.5){i_*}[rr] && (i_*)^*(V(\M)) \ar
|(0.5){\pi} [uur] \ar |(0.6){i_*}[rr] \ar@/^1pc/@{<.>}
|(.65){\Delta_{\bf h}}[uuuur]  &&V(\M)\ar |(0.5){\pi} [uur]
\ar@/^1pc/@{<.>} |(.65){\Delta_{\bf h}}[uuuur] && \\ \\%
V(\B)\ar |(0.5){i} [uu] \ar |(0.4){i_*}[rr] && i_*(V(\B)) \ar
|(0.5){i} [uu]  && (i_*)^*(V(\A)) \ar |(0.5){i} [uu] \ar
|(0.6){i_*}[rr] &&V(\A)\ar |(0.5){i} [uu]
&&%
}%
\end{equation*}
With a mild abuse of notation, identifying some bundles with their
immersions, we can focus our attention on the central part of the
diagram, synthetically represented by
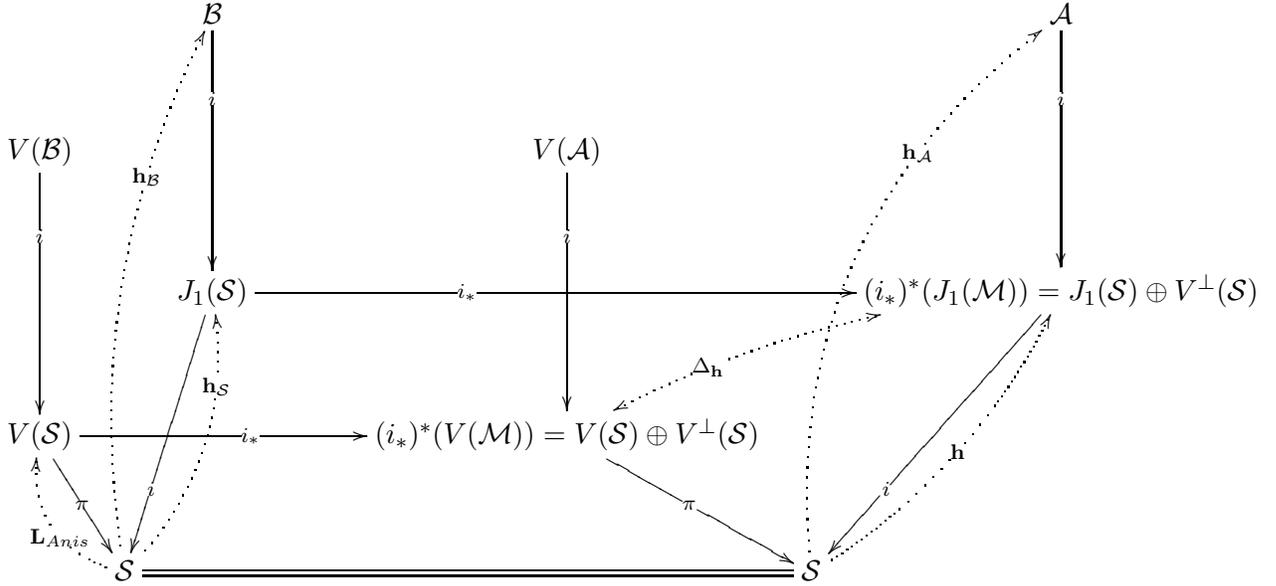
\begin{equation*}
\hskip-1.5truecm
\xymatrix @R=15pt @C=10pt{%
& & \B \ar |(0.3){i} [dddd] & & & & &  \A \ar |(0.3){i} [dddd] \\ \\%
V(\B) \ar |(0.3){i} [dddd] & & & & & V(\A) \ar |(0.3){i} [dddd] & & & &\\ \\%
& & \JS \ar |(0.7){i} [ddddl] \ar |(0.3){i_*}[rrrrr] & & & & & (i_*)^*(\J)=\JS\oplus V^{\perp}(\S)\ar |(0.7){i} [ddddl] &\\ \\%
V(\S) \ar |(0.4){i_*}[rrrrr] \ar |(0.5){\pi} [ddr] & & & & &
(i_*)^*(V(\M))=V(\S)\oplus V^{\perp}(\S) \ar |(0.5){\pi} [ddr]
\ar@/^1pc/@{<.>} |(.3){\Delta_{\bf h}}[uurr] & & &
\\ \\%
& \S \ar@{=}[rrrrr] \ar@/^-1.2pc/@{..>} |(.7){{\bf h}_{\S}}[uuuur]
\ar@/^1.5pc/@{..>} |(.7){{\bf h}_{\B}}[uuuuuuuur]
\ar@/^1.2pc/@{..>} |(.4){{\bf L}_{Anis}}[uul] & & & & & \S
\ar@/^-.9pc/@{..>} |(.5){\bf h}[uuuur] \ar@/^2.8pc/@{..>}
|(.7){{\bf h}_{\A}}[uuuuuuuur] & &
}%
\end{equation*}
Later on, with the same mild abuse of notation used in the
previous diagrams, every frame ${\bf h}$ having $\S, \A$ or $\B$
as index are intended such that, once restricted to the points of
$\S$, the frame has image in $\JS, \A$ or $\B$ respectively, and
than can be considered, naively speaking, a rest frame of the
corresponding constraint.

\section{Impulsive problem and constitutive characterization of
constrained systems}

The impulsive problem for a constrained system has a formulation
very similar to that of the impulsive problem for free system, but
the mechanical foundations of the two problems are very different.
Both consists in the determination of a uniquely determined right
velocity ${\bf p}_R$ (belonging to a suitable bundle)  once the
left velocity ${\bf p}_L$ (chosen in general as element of a
suitable bundle) is known. In both cases the mechanical law
governing the phenomenon is simply the action (\ref{ActionVJ})
such that ${\bf p}_R \= {\bf p}_L + {\bf I}$ where ${\bf I}$ is
the impulse acting on the system. The crucial difference relies in
the application of the principle of determinism of Classical
Mechanics.

In the case of free systems, both ${\bf p}_L$ and ${\bf p}_R$ are
not subject to any restriction. Then this also holds  for the
impulse ${\bf I}_{act}$, and any arbitrary rule assigning ${\bf
I}_{act}$ as function of ${\bf p}_L$ is admissible. We already saw
that in this case the principle of determinism entails that the
active impulse ${\bf I}_{act}({\bf p}_L)$ must be an a priori
known data of the problem.

In the case of constrained systems, ${\bf p}_L$ and ${\bf p}_R$
are subject to the restrictions given by the constraints.  Then
the so called reactive impulse ${\bf I}_{react}$  cannot be
assigned a priori as an arbitrary function of ${\bf p}_L$, but it
is an unknown of the problem, obeying the condition that, for
every admissible choice of ${\bf p}_L$, the right velocity ${\bf
p}_R$ satisfies the restriction of the constraints. Moreover, in
application of the principle of determination, the right velocity
must be univocally defined.

The most general form of the INL in case of constrained system
assumes then the form ${\bf p}_R = {\bf p}_L + {\bf I}_{act}({\bf
p}_L) + {\bf I}_{react}({\bf p}_L)$.

\begin{Definition}
An impulsive constitutive characterization for the constraints
acting on the system is a rule determining characteristics and
properties of the reactive impulse ${\bf I}_{react}$ in a way
assuring the respect of the constraints and of the principle of
determinism of Classical Mechanics.
\end{Definition}

\begin{Example} Let a system be subject only to a unilateral
positional constraint $\S$. Then for every left velocity ${\bf
p}_L \in \L_{\S}(\J)$ the impulsive problem can exists even in
absence of active impulse and consists in determining the right
velocity in the form ${\bf p}_R \= {\bf p}_L + {\bf
I}_{react}({\bf p}_L) \in \R_{\S}(\J)$. The constitutive
characterization of $\S$ consists in a rule that allows to
determine univocally ${\bf I}_{react}({\bf p}_L) \in
(i_*)^*(V(\M))$, and then ${\bf p}_R$, once ${\bf p}_L$ is known.
\end{Example}

Differently from the free systems, constrained systems can then
have an impulsive behavior even in absence of active impulses.
There are then two different classes of constitutive
characterization: the one modelling the impulsive reaction of the
constraint in absence of active impulses and the one modelling the
impulsive reaction of the constraint when the system is subject to
an active impulse. Due to the stiltedness of the concept of active
impulse, the first one will be the main, but not the only, focus
of the following analysis.

\begin{Example} Let a system be subject only to a permanent kinetic
constraint $\A$. In this case, the impulsive problem exists only
if an active impulse ${\bf I}_{act}({\bf p}_L) \in (i_*)^*(V(\M))$
is a priori assigned. Since both ${\bf p}_L, {\bf p}_R$ must be in
$\A$, then the total impulse ${\bf I}_{react}({\bf p}_L) + {\bf
I}_{act}({\bf p}_L) = {\bf p}_R - {\bf p}_L \in V(\A)$. Therefore,
in this case the constitutive characterization of $\A$ consists in
a rule that allows to determine ${\bf I}_{react}({\bf p}_L) \in
(i_*)^*(V(\M))$ once ${\bf p}_L$ and ${\bf I}_{act}({\bf p}_L)$
are known. Note however that, since in general ${\bf I}_{act}({\bf
p}_L) \in (i_*)^*(V(\M)) =  V(\A) \oplus V^{\perp}(\A)$, then
every possible constitutive characterization for $\A$ must obey
the condition $\PAO({\bf I}_{react}({\bf p}_L)) \= - \PAO({\bf
I}_{act}({\bf p}_L))$ and then  the constitutive characterization
is determined by a rule assigning $\PA({\bf I}_{react}({\bf
p}_L))$ as function of ${\bf p}_L$ and ${\bf I}_{act}({\bf p}_L)$.
\end{Example}

The wide generality of the definition of constitutive
characterization of constraints constitutes the fertile arena
where we can model different physical behaviors of the same
geometric constraint. For example, the two well known concepts of
ideality and friction find their logical foundation in the context
of constitutive characterization, and the same geometric
constraint can be ideal or not, frictionless or not. Vice versa,
the same principle inspiring a constitutive characterization can
determine different effective characterizations when applied to
geometrically different constraints, so that the ideality of
single or multiple constraints can be performed through different
rules.

\begin{Remark} Neglecting the forces acting on an impulsive mechanical
system is an usual assumption in the context of Impulsive
Mechanics. This seems only partially reasonable. In fact, if we
consider the reactive forces that act on the system in the very
short time interval of the impact phenomenon, their action is
efficiently modelled by assuming as instantaneous the impact and
introducing the reactive impulse ${\bf I}_{react}({\bf p}_L)$.
However, neglecting the influence of the active forces on the
behavior of an impulsive system seems less reasonable. For
instance, let us consider Ex.10 and Ex.11: it is an experimental
evidence that in the impact of the sphere/disk with the vertical
wall, the permanent kinetic constraint $\A$ can break, so that the
permanent rolling condition verified before the impact can be not
verified after the impact. Such an eventuality surely depends on
the magnitude of $\velort({\bf p}_L)$, but it is inconceivable
that it does not depend on the weight of the sphere/disk, that
tightens the contact between the sphere/disk and the horizontal
plane/line.

\smallskip

The efficacy of the geometric framework introduced above is
pointed out further by showing that the active forces acting on
the system can enter in the choice of the constitutive
characterization of the constraint.

The second jet--extension $\pi: \JJ \to \J$ of the bundle $\M$
(see \cite{MassPaga1991}) is the {\it absolute acceleration
space--time} of the system. It is the $(3n+1)$--dimensional affine
subbundle of the tangent bundle $T(\J)$ whose elements have the
form ${\bf p} \= \frac{\partial}{\partial t} + \dot{x}^i
\frac{\partial}{\partial x^i}+ a^i \frac{\partial}{\partial
\dot{x}^i}, \, i=1,\ldots,n$, so that $\JJ$ can be locally
described by jet--coordinates $(t, x^i,\dot{x}^i, \ddot{x}^i)$.

The affine jet--bundle $\JJ \to \J$ is modelled on the
$(3n+1)$--dimensional vector bundle $V(\J)\to \J$ of the vertical
vectors of $T(\J)$, that is the vectors that are tangent to the
fiber of $\pi:\J\to \M$. Using admissible coordinates, the
elements of $V(\J)$ have the form ${\bf Z} \=  Z^i
\frac{\partial}{\partial \dot{x}^i}, \, i=1,\ldots,n$, so that
$V(\J)$ too can be locally described by jet--coordinates $(t,
x^i,\dot{x}^i, \ddot{x}^i)$. The correspondence
$\frac{\partial}{\partial x^i} \leftrightarrow
\frac{\partial}{\partial \dot{x}^i}$ gives a natural isomorphism
$\Upsilon$ of the vertical spaces $V(\M)$ and $V(\J)$.

The assignment of the active forces acting on the system (see once
again \cite{MassPaga1991}) consists in the assignment of a section
$\Theta: \J \to V(\J)$, locally expressed by the functions $Z^i \=
Z^i(t,x^j,\dot{x}^j), \, i,j=1,\ldots,n$. Therefore, the knowledge
of the active forces acting on the (impulsive) system determines
the knowledge of a map $\Upsilon(\Theta({\bf p}))$. This is enough
to allow that the active forces (alternatively expressed in the
form $\Upsilon(\Theta({\bf p}))$ or in the local form $Z^i \=
Z^i({\bf p})$) can be taken into account in the choice of the
constitutive characterization ${\bf I}_{react} \= {\bf
I}_{react}({\bf p}_L, \Upsilon(\Theta({\bf p}_L)))$. \end{Remark}

In this section, we describe some of the most common constitutive
characterizations for the main classes of constraints
geometrically distinguished as above, mainly in absence of active
impulse.

\subsection{Ideality}

The standard requirement of ideality for a constraint acting on an
system is that the reaction does not perform power or work. This
cannot be required in an impulsive phenomenon, that is
instantaneous and without variation of position.

In presence of active impulse, the most natural requirement of
ideality can repeat the standard arguments of non impulsive
Classical Mechanics, for instance requiring the absence of a
``tangent'' component of the reactive impulse.

\begin{Example} Let $\S$ a bilateral positional constraint and ${\bf
I}_{act}({\bf p}_L) \in (i_*)^*(V(\M)) =  V(\S) \oplus
V^{\perp}(\S)$ with $\P^{\perp}({\bf I}_{act}({\bf p}_L)) \neq 0$.
Note that this is a case of system subject to a permanent
positional constraint whose presence could reduce the dimension of
the space--time configuration of the system  but such that the
reduction is not admissible for the nature itself of the impulsive
problem.

With a slight abuse of notation and recalling Ex.13, a possible
ideal criterion is given in this case by the conditions $\PVO({\bf
I}_{react}({\bf p}_L)) \= - \PVO({\bf I}_{act}({\bf p}_L))$ and
$\PV({\bf I}_{react}({\bf p}_L)) \= 0$.

The analogous choice $\P^{\|}_{\A}({\bf I}_{react}({\bf p}_L)) \=
0$ can express the ideality of a permanent kinetic constraint $\A$
in presence of a generic active impulse ${\bf I}_{act}({\bf p}_L)
\in (i_*)^*(V(\M)) =  V(\A) \oplus V^{\perp}(\A)$. \end{Example}

In absence of active impulses, the most natural requirement of
ideality consists in the preservation of the kinetic energy of the
system before and after the impulsive action of the constraint.
Let us show that, as we previously sketched in Section 1, a naive
formulation of this requirement determines obvious inconsistencies
of the approach, even for very simple mechanical system.

\begin{Example} We consider a particular impulsive motion of a rod of length
$2L$ and mass $M$ moving in a halfplane. The space--time bundle
can be described by coordinates $(t,x,y,\vth)$ where $(x,y)$ are
the cartesian coordinates of the center of the rod and $\vth$ is
its orientation. The vertical metric is expressed by the mass
matrix $g_{ij} = diag(M,M,A)$ with $A=\frac13ML^2$. The unilateral
constraint $\S$ can be locally described by the condition $y-L\sin
\vth=0$. We suppose that the rod, moving downward in vertical
position with vertical velocity $-\dot{y}_0$ respect to
constraint, impacts with the constraint and rebounds in vertical
position with vertical velocity $\dot{y}_0$. Using admissible
coordinates we suppose that the impact happens in the point
$(t_0,x_0,L,\pi/2)$ with absolute velocities
\begin{eqnarray*}
\begin{array}{l}
{\bf p}_L \= \dfrac{\partial}{\partial t} \, - \, \dot{y}_0 \,
\dfrac{\partial}{\partial y} \\
{\bf p}_R \= \dfrac{\partial}{\partial t} \, + \, \dot{y}_0 \,
\dfrac{\partial}{\partial y} \, .
\end{array}
\end{eqnarray*}
Intuitively, the impact seems to have an ideal behavior, since the
kinetic energy seems preserved in the impact. However, as we
already saw, this is a meaningless statement until we do not
specify the frame of reference where the kinetic energy is
preserved. Note in fact that, if we introduce the three frames of
reference:
\begin{eqnarray*}
\begin{array}{lll}
{\bf h}_0 \= \dfrac{\partial}{\partial t} \, + \, H_x(t,x,y) \,
\dfrac{\partial}{\partial x} \, , \\ \\
{\bf h}_1 \= \dfrac{\partial}{\partial t} \, + \, H_x(t,x,y) \,
\dfrac{\partial}{\partial x} \, + \, \dot{y}_0 \,
\dfrac{\partial}{\partial y} \, , \\ \\
{\bf h}_2 \= \dfrac{\partial}{\partial t} \, + \, H_x(t,x,y) \,
\dfrac{\partial}{\partial x} \, - \, \dot{y}_0 \,
\dfrac{\partial}{\partial y} \, ,
\end{array}
\end{eqnarray*}
the kinetic energy of the system in the impact is preserved for
${\bf h}_0$, decreases to zero for ${\bf h}_1$ and even increases
from zero to a positive value for ${\bf h}_2$. This is (obviously)
due to the facts that ${\bf h}_0$ is a rest frame for the
constraint, ${\bf h}_1$ is a ``comoving'' frame of the rod after
the impact and ${\bf h}_2$ is a ``comoving'' frame of the rod
before the impact. \end{Example}

The example above shows that, of course, preservation of kinetic
energy cannot be required for all the frames of reference of the
system. On the other side, requiring the preservation in a single
frame conflicts with the basic requirement that a physical
property of a system must be independent on the frame of
reference.

It is a known result (see \cite{Pasquero2005uni,Pasquero2006})
that the requirement of preservation of the kinetic energy for all
the frames of reference in the class of the rest frames of the
constraints is sufficient (in absence of active impulses) to
determine a satisfactory constitutive characterization for some
significant classes of constraints, such as unilateral constraints
of codimension $1$ (\cite{Pasquero2005uni}), also in presence of
permanent and/or instantaneous kinetic constraints
(\cite{Pasquero2006}). In these cases, roughly speaking, the
impulsive reaction is determined by the orthogonal component
$\velort({\bf p}_L)$ of the left--velocity in the form ${\bf
I}_{react}({\bf p}_L) = -2 \velort({\bf p}_L)$, and the impulsive
behavior of the system is essentially a reflection with respect to
the ``orthogonal direction'' determined by the constraint.

\begin{Remark} The impact of Ex.15 is really ideal since the kinetic energy
is preserved for all the rest frames of the constraint $y-L\sin
\vth=0$ (and in particular for ${\bf h}_0$). \end{Remark}

It is also known that the only preservation of kinetic energy is
not sufficient to determine univocally the constitutive
characterization for some classes of constraints (such as
positional constraints of codimension greater than $1$ and
 multiple unilateral constraints).
Nevertheless, the ``reflection'' characterization ${\bf
I}_{react}({\bf p}_L) = -2 \velort({\bf p}_L)$ could be (such as
in the case of positional constraints of codimension greater than
$1$. See once again \cite{Pasquero2005uni}) or at least could
suggest (such as in the case of multiple unilateral constraints.
See e.g. \cite{Pasquero2017QAM}) physically meaningful ideal
constitutive characterizations.

\begin{Remark} Of course the requirement of preservation of the kinetic
energy is in general not admissible in presence of active
impulses, as clearly shown by Ex.14. \end{Remark}

Needless to say, the importance of the ideal (and in particular
the ``reflection'') characterization relies not only in its
structural simplicity (it involves only the geometric structures
determined by the constraints) and its effective applicability to
several meaningful systems, but also because it constitutes the
starting point to analyze non ideal behaviors of constraints.

\subsection{Frictionless non--ideality}

The ideal requirement of preservation of kinetic energy in absence
of active impulse suggests the possible non--ideal
characterization where the kinetic energy is partially or totally
lost in a non--elastic impact. Once again, the naive idea of a
fixed percentage of loss of kinetic energy, expressed for example
by an energetic restitution coefficient $\varepsilon_{\K} \in
[0,1)$, gives rise to inconsistencies.

\begin{Example} The same rod of Ex.15 vertically falls but does not
rebound on the constraint, so that its time evolution after the
impact is given by the motion $\gamma(t) = (t, x_0, L, \pi/2)$ and
${\bf p}_R = \frac{\partial}{\partial t}$. Taking into account
that ${\bf h}_0 = \frac{\partial}{\partial t} \, + \, H_x \,
\frac{\partial}{\partial x} \in \H_{\S}$ for every $H_x$, with
obvious notation we have that
\begin{eqnarray*}
\begin{array}{l}
\varepsilon_{\K} \= \dfrac{(\K_{{\bf h}_0})_R}{(\K_{{\bf h}_0})_L}
\= \dfrac{M (H_x)^2}{M (H_x)^2 + M (\dot{y}_0)^2} \, ,
\end{array}
\end{eqnarray*}
that explicitly depends on $H_x$. \end{Example}

The example above shows that the percentage of loss of kinetic
energy in a non--elastic impact is not frame invariant even if we
restrict the assignment of $\varepsilon_{\K}$ to the class of rest
frames of the constraint. A frame invariant formulation for the
restitution coefficient can be obtained by considering the
percentage of ``reflection'' of the orthogonal velocity in the
impact: a (Newtonian) coefficient $\varepsilon^{\perp} \in [0,1)$
such that
\begin{eqnarray*}
{\bf I}_{react}({\bf p}_L) = -(1+\varepsilon^{\perp}) \,
\velort({\bf p}_L)
\end{eqnarray*}
has a clear invariant (with respect to the rest frames) meaning.

The particular case $\varepsilon^{\perp} = 0$ naturally gives the
non--ideal {\it totally inelastic} characterization, defined as
\begin{eqnarray*}
\begin{array}{rccll}
TotIn: & \L_{\S}(\J) & \to & (i_*)^*(V(\M)) &\quad \mbox{s.t.}\\
&{\bf p}_L & \rightsquigarrow & {\bf I}_{TotIn}( {\bf p}_L )\= -\,
\velort( {\bf p}_L) \, .
\end{array}
\end{eqnarray*}
The evolution equation assumes the form
\begin{eqnarray*}
\begin{array}{ccll}
\L_{\S}(\J) & \to & \R_{\S}(\J) &\quad \mbox{s.t.}\\
{\bf p}_L  & \rightsquigarrow & {\bf p}_R \= {\bf p}_L -\, \velort( {\bf p}_L) \= \PJ({\bf p}_L)  \, .\\
\end{array}
\end{eqnarray*}
Note that, in this case, there exists a subclass of the rest
frames $\H_{\S}$ of $\S$ formed by all those frames such that
${\bf h}(\pi({\bf p}_L)) = \PJ({\bf p}_L)$ for which the system
stops after the impact and then have null kinetic energy. Of
course this property does not hold for all the frames of
$\H_{\S}$.

\subsection{Friction}

The naive idea of impulsive constraint with friction can be
expressed by the condition $\PV({\bf I}_{react}({\bf p}_L)) \neq
0$. Once again, due to the wide variety of possible constraints
acting on impulsive systems, there are several different way to
apply the naive idea sketched above to the various constraints.

Given a system subject to a bilateral positional constraint $\S$
that is not included in the construction of the space--time bundle
$\M$ (resp. a system subject to a kinetic permanent constraint
$\A$) and subject to an active impulse that is not ``tangent'' to
$\S$, that is ${\bf I}_{act} \in (i_*)^*(V(\M))$ but ${\bf
I}_{act} \notin V(\S)$ (resp. ${\bf I}_{act} \notin V(\A)$), the
standard arguments of Classical (non Impulsive) Mechanics can be
applied replacing active and reactive forces with active and
reactive impulses: the projection operators $\PV, \PVO$ determine
the tangent and parallel components of the active impulse and a
constitutive characterization (for example of Coulomb type) of the
reactive impulse can be assigned as functions of these components.

\begin{Example} A massive particle moving on a horizontal floor is
subject to an active impulse ${\bf I}_{act}$ that is not parallel
to the floor. The $4$--dimensional space--time $\M$ can be
described by the coordinates $(t,x,y,z)$, where $(x,y,z)$ are the
coordinates of the particle, together with the permanent
positional constraint $\S = \{z=0\}$. The active impulse ${\bf
I}_{act} \in (i_*)^*(V(\M))$ but ${\bf I}_{act} \notin V(\S)$.

Once again, since both ${\bf p}_L, {\bf p}_R \in \JS$, the total
impulse ${\bf I}_{act}({\bf p}_L) + {\bf I}_{react}({\bf p}_L) =
{\bf p}_R - {\bf p}_L \in V(\S)$. Then the condition $\PVO({\bf
I}_{react})) \= - \, \PVO({\bf I}_{act})$ is mandatory in order to
respect the constraint. We can then introduce a constitutive
characterization with friction for $\S$ by assigning $\PV({\bf
I}_{react}({\bf p}_L))$ as a non--null function of $({\bf p}_L,
\PV({\bf I}_{act}({\bf p}_L)), \PVO({\bf I}_{act}({\bf p}_L)))$.
\end{Example}

\begin{Example} A massive particle moving in a $3$--dimensional
euclidean space is subject to a permanent  kinetic constraint $\A$
and to an active impulse ${\bf I}_{act}({\bf p}_L) \in
(i_*)^*(V(\M)), {\bf I}_{act}({\bf p}_L) \notin V(\A)$. Since the
total impulse ${\bf I}_{act}({\bf p}_L) + {\bf I}_{react}({\bf
p}_L) = {\bf p}_R - {\bf p}_L \in V(\A)$, then
$\P^{\perp}_{\A}({\bf I}_{react}({\bf p}_L)) \= - \,
\P^{\perp}_{\A}({\bf I}_{act}({\bf p}_L))$ and a constitutive
characterization of $\A$ can be assigned determining
$\P^{\|}_{\A}({\bf I}_{react}({\bf p}_L))$ as a non--null function
of $({\bf p}_L, \P^{\|}_{\A}({\bf I}_{act}({\bf p}_L)),
\P^{\perp}_{\A}({\bf I}_{act}({\bf p}_L)))$. \end{Example}

For impulsive systems that are not subject to active impulses, the
analogies between classical and impulsive description of friction
is in general not possible. This is mainly due to geometrical
reasons.

Let $\S$ be a unilateral positional constraint: the condition
$\PV({\bf I}_{react}({\bf p}_L)) \neq 0$ can be taken as starting
point to define a constitutive characterization with friction for
$\S$, but the effective applicability of this idea must, in
general, take into account the lack of prior specific directions
in $V(\S)$. The system of Ex.3 gives the simplest example of this
situation.

The usual (and sometimes unaware) way to select a prior direction
in $V(\S)$ consists in the assignment of a frame of reference
${\bf h}_{\S} \in \H_{\S}$ thought of as ``the'' rest frame of the
constraint. In fact, in this case, we can consider the tangent
component  $\vel^{\|}_{{\bf h}_{\S}}({\bf p}_L) = \PJ({\bf p}_L) -
{\bf h}_{\S} = \PV({\bf p}_L - {\bf h}_{\S})$ of the velocity, and
we can choose $\PV({\bf I}_{react}({\bf p}_L)) \in
Lin\{\vel^{\|}_{{\bf h}_{\S}}({\bf p}_L)\}$. The constitutive
characterization can then be assigned in the form
\begin{eqnarray}
\left\{
\begin{array}{l}
\PVO({\bf I}_{react}({\bf p}_L)) \= \alpha({\bf p}_L) \,
\vel^{\perp}_{{\bf h}_{\S}}({\bf p}_L)\\ \\
\PV({\bf I}_{react}({\bf p}_L)) \= \beta({\bf p}_L) \,
\vel^{\|}_{{\bf h}_{\S}}({\bf p}_L)
\end{array}
\right.
\end{eqnarray}
for suitable choices of the functions $\alpha({\bf p}_L),
\beta({\bf p}_L)$.

\begin{Remark}
The presence of a non--null frictional component $\PV({\bf
I}_{react}({\bf p}_L)) \neq 0$ does not imply the non--ideality of
the characterization of the constraint. The ideal characterization
of the set of constraints comprised of a unilateral positional
constraint $\S$ and an instantaneous kinetic constraint $\B$, due
to the splitting $V(\M) \= V(\B) \oplus V^{\perp}(\B) \oplus
V^{\perp}(\S)$ (see \cite{Pasquero2006}) gives an example of ideal
constitutive characterization with friction.
\end{Remark}

\subsection{Breakability}

The naive idea of systems subject to a breakable impulsive
constraint is that of systems whose behavior respects the
constraint condition before the impulsive phenomenon and does not
respect the constraint condition after the impulsive phenomenon.
There are many physical examples of this possibility: for
instance, a bullet that can bounce on or perforate a wall, or a
billiard ball that rolls on the table before the impact with the
cushion and that slides after the impact.

In order to give a geometrical description of breakability of a
positional constraint $\S$, we introduce a generalization of the
spaces $\L_{\S}(\J), \R_{\S}(\J)$ of Section 2.1: we say that $\S$
is breakable if $\L_{\S}(\J) \cap \R_{\S}(\J) \ne \emptyset$. This
simply means that there are some admissible left velocities that
can also be admissible right velocities.

\begin{Example} Let $\S$ be a positional constraint of codimension
$1$, let $\L_{\S}(\J) \= \{{\bf p}_L \, | \, \Phi(\velortS({\bf
p}_L), {\bf U}^{\perp}) < 0 \}$ and let the constitutive
characterization of $\S$ be such that ${\bf I}_{react}({\bf p}_L)
\= - \, \lambda_{\Xi}({\bf p}_L) \, \velortS({\bf p}_L)$ with
\begin{eqnarray*}
\lambda{_\Xi}({\bf p}_L) \= 2 \, \dfrac{\Xi^2}{\Xi^2 +   \|
\velortS({\bf p}_L) \|^2} \, ,  \qquad \Xi>0 \, .
\end{eqnarray*}
In this case, $\S$ is ``almost'' elastic for $\| \velortS({\bf
p}_L) \| \ll 1$, it is anelastic for $\| \velortS({\bf p}_L) \| <
\Xi$, it is totally anelastic for $\| \velortS({\bf p}_L) \| \=
\Xi$, it is broken for $\| \velortS({\bf p}_L) \| > \Xi$. In this
case the value $\Xi$ represents the ``breakability threshold'' of
$\S$.

A similar characterization gives, for example, a model for the
impulsive behavior of a bulletproof glass. \end{Example}

\begin{Example} Let $\S$ and $\L_{\S}(\J)$ be as above and let the constitutive
characterization of $\S$ be such that ${\bf I}_{react}({\bf p}_L)
\= - \, \lambda_{\Xi}({\bf p}_L) \, \velortS({\bf p}_L)$ with
\begin{eqnarray*}
\lambda{_\Xi}({\bf p}_L) \= 2 \, \dfrac{\| \velortS({\bf p}_L)
\|^2}{\Xi^2 +   \| \velortS({\bf p}_L) \|^2} \, ,  \qquad \Xi>0 \,
.
\end{eqnarray*}
In this case, $\S$ is ``almost'' elastic for $\| \velortS({\bf
p}_L) \| \gg 1$,  it is totally anelastic for $\| \velortS({\bf
p}_L) \| \= \Xi$, it is broken for $\| \velortS({\bf p}_L) \| <
\Xi$. Once again the value $\Xi$ represents the ``breakability
threshold'' of $\S$.

A similar characterization gives, for example, a model for the
impulsive behavior of the surface of a non--newtonian fluid.
\end{Example}

An impulsive action (for instance, an active impulse or the impact
with a unilateral positional constraint $\S$) can break the action
of a kinetic ``permanent'' constraint $\A$. In this case the
breakability of $\A$ can be modelled by assuming $A$ or
$\L_{\S}(\A)$ as space of admissible left velocities and $\J$ or
$\R_{\S}(\J)$ as space of admissible right velocities.

\begin{Example} The disk of Ex.11 rolling without sliding on the
horizontal line and impacting with the vertical wall is subject to
the following constitutive characterization:
\begin{eqnarray*}
{\bf I}_{react}({\bf p}_L) \= \left\{
\begin{array}{lcl}
-(1+\varepsilon_1) \, \velort_{\JS\cap\A}({\bf p}_L) \quad & {\rm
if} & \quad  \|\velortS({\bf p}) \| \le \Xi\\ \\
-(1+\varepsilon_2) \, \velortS({\bf p}_L) \quad & {\rm if } &
\quad \|\velortS({\bf p}) \| > \Xi
\end{array}
\right.
\end{eqnarray*}
where $\varepsilon_1, \varepsilon_2 \in [0,1]$ and $\Xi \ge 0$. Of
course, the constants $\varepsilon_1, \varepsilon_2$ represent the
restitution coefficient of the contact disk/wall and $\Xi$
represents the breakability threshold of the pure rolling kinetic
constraint. Note moreover that, since the computation of
$\velortS({\bf p}_L)$ involves the mass $M$ of the disk, for
suitable choices of $\Xi$ a similar constitutive characterization
can embody the weight force acting on the disk and tightening the
contact between disk and horizontal line. \end{Example}

\section{Conclusions}

Pursuing the aim of establishing a correct setup where the main
concepts of Impulsive Mechanics of constrained system can be
expressed in agreement with the axioms and the invariance
requirements of Classical Mechanics, we showed that the geometric
environment given by the space--time configuration bundle $\pi_t:
\M \to \Euclideo$, its jet extensions and their subbundles allows:
\begin{itemize}
\item a meaningful embodiment of the absolute time and absolute
space axioms;

\item the coordinate invariance of concepts, physical quantities
and relations (granted by the standard techniques of Differential
Geometry);

\item a very general definition of frame of reference, and a
resulting ease of checking when frames are or not involved in the
definition of concepts, physical quantities and relations;

\item an invariant and causally correct formulation of the
Integrated or Impulsive Newton Law for free (unconstrained)
systems;

\item an invariant and causally correct formulation of the INL for
constrained systems that, through the concept of constitutive
characterization, respects the determinism of Classical Mechanics;

\item a clear distinction of the different classes of constraints
depending on their geometric properties;

\item a clear distinction of the different behavior of
constraints, depending on their constitutive characterization.

\end{itemize}
In particular, we obtained:
\begin{itemize}
\item the evidence that the geometric environments given by the
configuration space $\Q$ or its product bundle $\Reali \times \Q$
and $\Euclideo \times \Q$ are not appropriate to express the frame
invariance of concepts, physical quantities and relations of
Impulsive Mechanics, because of their intrinsic selection of a
frame of reference;

\item the evidence that the pure rolling constraint, despite the
theoretical possibility of being expressed in a frame invariant
way, can be operationally introduced only when a frame is chosen;

\item the evidence that the concept of friction can be introduced
in the constitutive characterization of a unilateral positional
constraint only if a rest frame of the constraint is chosen.

\end{itemize}
Along the paper, we also pointed out some unusual but meaningful
concepts and behaviors of impulsive systems:
\begin{itemize}

\item the concept of unilateral kinetic (permanent or
instantaneous) constraint;

\item the concept of breakable impulsive constraint;

\item the theoretical possibility that the active forces acting on
the system are involved in the definition of the impulsive
constitutive characterization of an impulsive constraint.

\end{itemize}



\end{document}